\documentclass{article}

\usepackage{arxiv}

\usepackage[utf8]{inputenc} 
\usepackage[T1]{fontenc}    
\usepackage{hyperref}       
\usepackage{url}            
\usepackage{booktabs}       
\usepackage{amsfonts}       
\usepackage{nicefrac}       
\usepackage{microtype}      
\usepackage{lipsum}
\usepackage{bm}
\usepackage{amsmath}
\usepackage{graphicx}
\usepackage{subfigure}
\usepackage{textcomp}
\usepackage{siunitx} 
\graphicspath{{./imgs/}}

\usepackage{framed ,color } 
\usepackage{pgfplots}
\pgfplotsset{compat=newest}
\pgfplotsset{plot coordinates/math parser=false}
\newlength\figureheight
\newlength\figurewidth

\title{Long short-term memory networks for proton dose calculation in highly heterogeneous tissues}

\author{
  Ahmad Neishabouri \\
  Department of Medical Physics in Radiation Oncology\\
  German Cancer Research Center (DKFZ)\\
  Im Neuenheimer Feld 280, D-69120 \\
  Heidelberg, Germany\\
  \texttt{a.neishabouri@dkfz-heidelberg.de} \\
   \And
 Niklas Wahl \\
  Department of Medical Physics in Radiation Oncology\\
  German Cancer Research Center (DKFZ)\\
  Im Neuenheimer Feld 280, D-69120 \\
  Heidelberg, Germany\\
  \texttt{n.wahl@dkfz-heidelberg.de} \\
  \AND
   Ulrich K\"othe\\
   Visual Learning Lab\\
   Interdisciplinary Center for Scientific Computing (IWR)\\
   University of Heidelberg\\
   Im Neuenheimer Feld 205 D-69120\\
   Heidelberg, Germany\\
   \texttt{ulrich.koethe@iwr.uni-heidelberg.de} \\
   \And
 Mark Bangert \\
  Department of Medical Physics in Radiation Oncology\\
  German Cancer Research Center (DKFZ)\\
  Im Neuenheimer Feld 280, D-69120 \\
  Heidelberg, Germany\\
  \texttt{m.bangert@dkfz-heidelberg.de} \\   
}

\begin{document}
\maketitle

\begin{abstract}
A novel dose calculation approach was designed based on the application of LSTM network that processes the 3D patient/phantom geometry as a sequence of 2D computed tomography input slices yielding a corresponding sequence of 2D slices that forms the respective 3D dose distribution. LSTM networks can propagate information effectively in one direction, resulting in a model that can properly imitate the mechanisms of proton interaction in matter. The study is centered on predicting dose on a single pencil beam level, avoiding the averaging effects in treatment plans comprised of thousands pencil beams. Moreover, such approach allows straightforward integration into today's treatment planning systems' inverse planning optimization process. The ground truth training data was prepared with Monte Carlo simulations for both phantom and patient studies by simulating different pencil beams impinging from random gantry angles through the patient geometry. For model training, 10'000 Monte Carlo simulations were prepared for the phantom study, and 4'000 simulations were prepared for the patient study. The trained LSTM model was able to achieve a 99.29 \% gamma-index pass rate ([0.5\%, 1 mm]) accuracy on the set-aside test set for the phantom study, and a 99.33 \% gamma-index pass rate ([0.5\%, 2 mm]) for the set-aside test set for the patient study. These results were achieved for each pencil beam in 6-23 ms. The average Monte Carlo simulation run-time using Topas was 1160 s. The generalization of the model was verified by testing for 5 previously unseen lung cancer patients. LSTM networks are well suited for proton therapy dose calculation tasks. However, further work needs to be performed to generalize the proposed approach to clinical applications, primarily to be implemented for various energies, patient sites, and CT resolutions/scanners.	
\end{abstract}


\section{Introduction}

The spatial calculation of the radiation dose within the patient's body is a central component of computer-aided treatment planning in the general radiotherapy chain. Thereby, accuracy is key - only a precise dose estimate enables a meaningful, patient-specific assessment of the treatment plan before the onset of therapy \cite{Newhauser2005, Newhauser2007, Bauer2014, Mein2019}.

At the same time, requirements regarding the dose calculation speed keep rising. Real-time dose calculation for adaptive radiotherapy (ultimately during treatment) \cite{Mein2018, Jia2012, Wang2017}, massively repeated dose calculation for uncertainty quantification \cite{Unkelbach2008, Kraan2013, Park2013, Bangert_2013, Wahl2017}, and complex simulations for biological effectiveness \cite{Mairani2013, Wieser_2017} are still too time-consuming for widespread clinical application.

For particle therapy, the trade-off between dose calculation speed and accuracy is defined by pencil beam algorithms on the one end and Monte Carlo algorithms on the other end. While pencil beam algorithms provide faster dose estimates, Monte Carlo algorithms require a higher computational load \cite{Fippel2004, Szymanowski2002}. At the same time, however, Monte Carlo algorithms clearly outperform pencil beam algorithms regarding accurarcy in complex geometries \cite{Schaffner1999, Soukup2005, Taylor2017}.

Currently, state-of-the-art deep learning (DL) technology is making an impact at various stages in radiotherapy. This process is most notably in classical machine learning domains such as outcome analysis and image processing \cite{gulliford2004use, chen2018u, nie2016estimating, bahrami20177t,gabrys2018}. Academic studies investigating deep learning for dose calculation are limited, and they primarily investigate the feasibility of deep learning methods in photon therapy \cite{Nguyen2019, Kontaxis_2020,pmlr-v85-mahmood18a, Kearney_2018}. Furthermore, considerations are restricted on training a \num{2}{D}/\num{3}{D} model (e.g. U-Net \cite{Ronneberger2015}) on the accumulative dose distribution extracted from prior-planned patient data, which poses problems for an application within inverse planning and seamless integration into existing workflows. This holds also true for recent work demonstrating the feasibility of improving protons dose calculation accuracy from the pencil beam algorithms to the level of Monte Carlo simulations, by learning from the prior-planned patient plans \cite{Wu2020}.

In this manuscript, we introduce a novel dose calculation approach for proton therapy based on the application of long short-term memory (LSTM) networks \cite{Hochreiter1997}, in an attempt to mimic the physics of proton interactions with matter in a single pencil beam level. We restrict this study to a minimal number of parameter dependence, and establish an end-to-end model that predicts the dose distribution based on the input CT. Therefore, the \num{3}{D} proton dose distribution of a pencil beam within the patient is understood as a sequence of two-dimensional dose slices along the beam direction. 

LSTM networks, unlike conventional feed-forward networks, have a hidden inner state enabling efficient processing of sequences of data and effective propagation of information along the sequence \cite{Sutskever_2014}. Currently, LSTM networks are applied highly successful for time-series data, e.\,g.\ stemming from speech or video \cite{Graves_2014,Donahue, ng2015short, kay2017kinetics}. To the best of our knowledge, this is the first work to exploit ANNs, and specifically LSTM networks, to perform proton dose calculation.

The designed approach will be further motivated in the following section \ref{sec_mNm} along with details on our LSTM architecture and training process. Section \ref{sec_results} presents results from a dose calculation accuracy study on model geometries and real-world lung patient cases. The limitations of our study and general opportunities provided through LSTM network proton dose calculations are discussed in section \ref{sec_discussion}, section \ref{sec_conclusion} concludes the paper.

\section{Material and methods}
\label{sec_mNm}The elementary task underlying the dose calculation for an entire intensity modulated proton therapy treatment is the calculation of the dose of a single proton pencil beam. In this context, a pencil beam denotes a bunch of protons leaving the treatment nozzle with a reasonably confined momentum distribution, as determined by the beam shaping devices.

Consequently, our study focuses on considerations for individual pencil beams. This reduction was chosen to study the fundamental characteristics of LSTM network-based dose calculations without averaging effects in treatment plans comprised of thousands of pencil beams which may conceal important aspects regarding the accuracy of the physical dose deposition.

\subsection{Conventional proton dose calculation}

With conventional dose calculation approaches, the \num{3}{D} dose distribution of a single pencil beam within the patient body $\boldsymbol{\mathcal{D}}$ is a function of the initial phase space (i.,\,e., the initial position and momentum distribution) of the particles $\boldsymbol{\mathcal{P}}$ and the \num{3}{D} patient geometry $\boldsymbol{\mathcal{G}}$.

\begin{align}
	\boldsymbol{\mathcal{D}} = f(\boldsymbol{\mathcal{P}}, \boldsymbol{\mathcal{G}})
\end{align}

The patient geometry is usually determined with a computed tomography (CT) scan where the Hounsfield units (HU) get translated into material composition distributions for Monte Carlo algorithms with custom calibration curves. Based on samples from the initial phase space of the particles, Monte Carlo algorithms simulate the path of individual protons and the associated energy deposition within the patient, as determined by its interactions with the patient geometry. The final dose distribution is then given by the sum of the deposited energy of all simulated particles. While this approach allows for highly precise dose estimates with sufficient histories being simulated, even in challenging geometries, the repeated simulation of individual particles is very time consuming.

In our study, the Monte Carlo dose calculations were carried out with the Topas (TOol for PArticle Simulation) wrapper \cite{Perl2012} for Geant4 \cite{Agostinelli:2002hh}. The initial particle energy was \SI{104.25}{\mega\electronvolt} for all simulations providing a reasonable trade-off between a meaningful penetration depth and acceptable dose calculation as well as LSTM network training run-times during prototyping.

\subsection{Neural networks for proton dose calculation}

In order to train a neural network for proton dose calculations, it is necessary to learn a mapping from the \num{3}{D} patient geometry $\boldsymbol{\mathcal{G}}$ and the initial particle phase space $\boldsymbol{\mathcal{P}}$ to the \num{3}{D} dose distribution $\boldsymbol{\mathcal{D}}$, as laid out in the previous section. In the following two subsections, we are going to explain (1) our parameterization of the proton dose calculation problem for a neural network and (2) the rationale underlying our network architecture.

\subsubsection{Problem parameterization}
\label{sec:parameterization}

In order to minimize the complexity of the training process for the neural network, we restrict the transformation to be learned for dose calculation to a single initial energy. Without the loss of generality (we can simply train one network per initial energy), this effectively reduces the space of possible dose calculation scenarios for the network and enables a denser sampling of the space of possible patient geometries and dose distributions.

\begin{figure}[htb]
	\begin{center}
		\centering
		\subfigure[]{\includegraphics[width=0.6\linewidth]{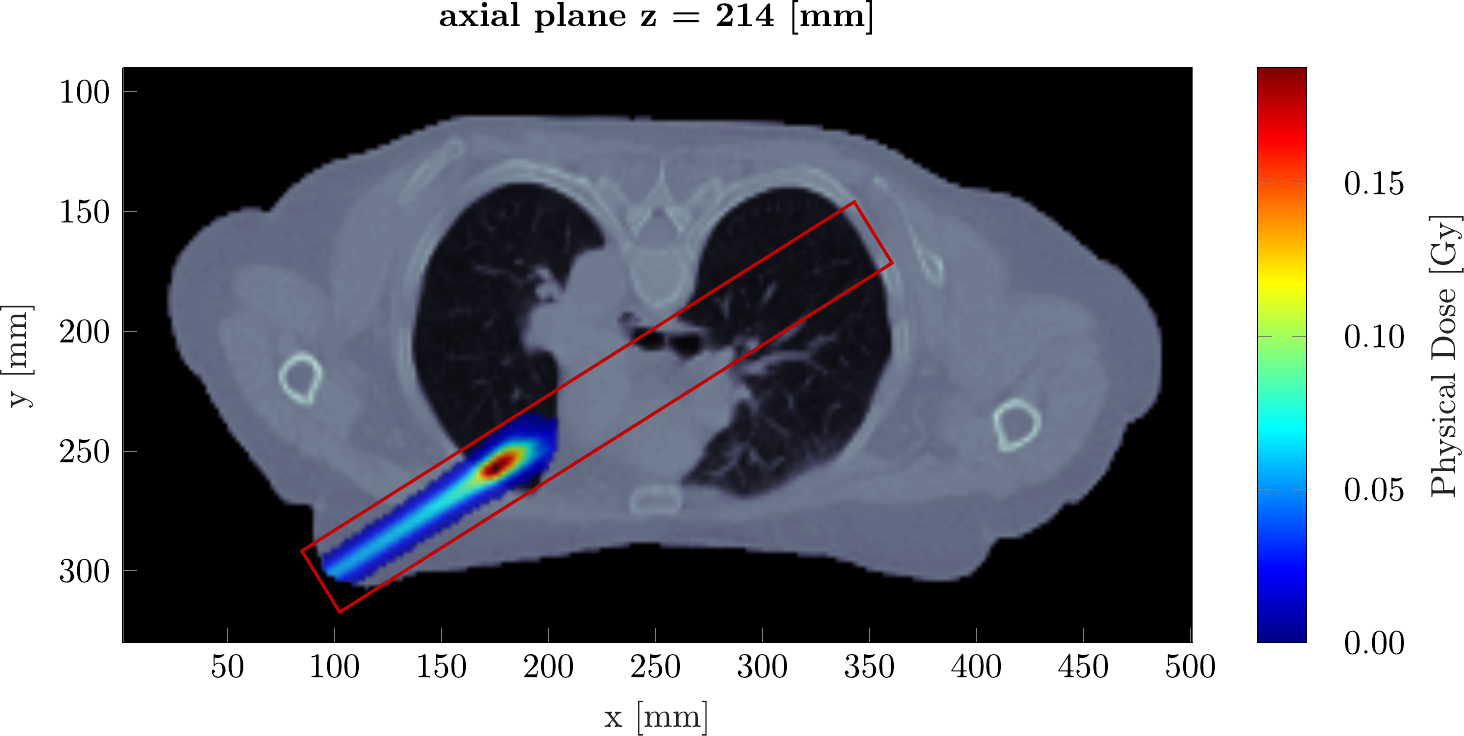}\label{fig:MandM_1a}} \\
		\subfigure[]{\includegraphics[width=0.63\linewidth]{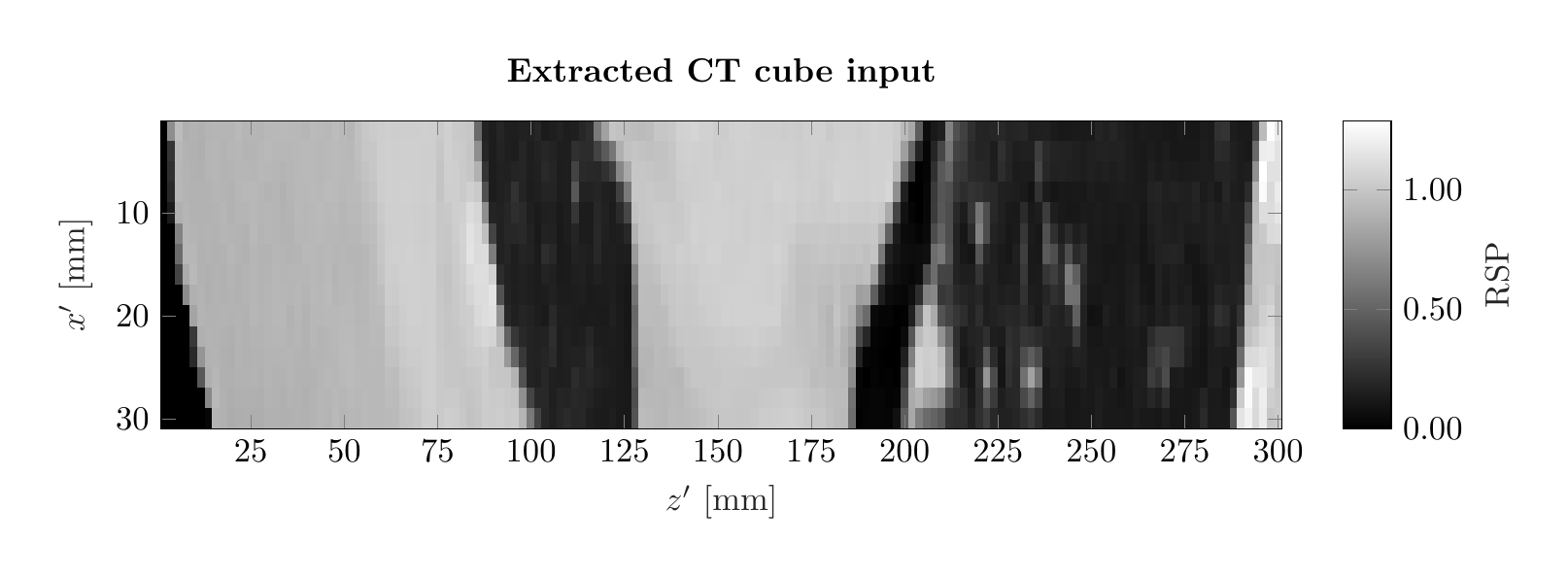}\label{fig:MandM_1b}} \\
		\subfigure[]{\includegraphics[width=0.63\linewidth]{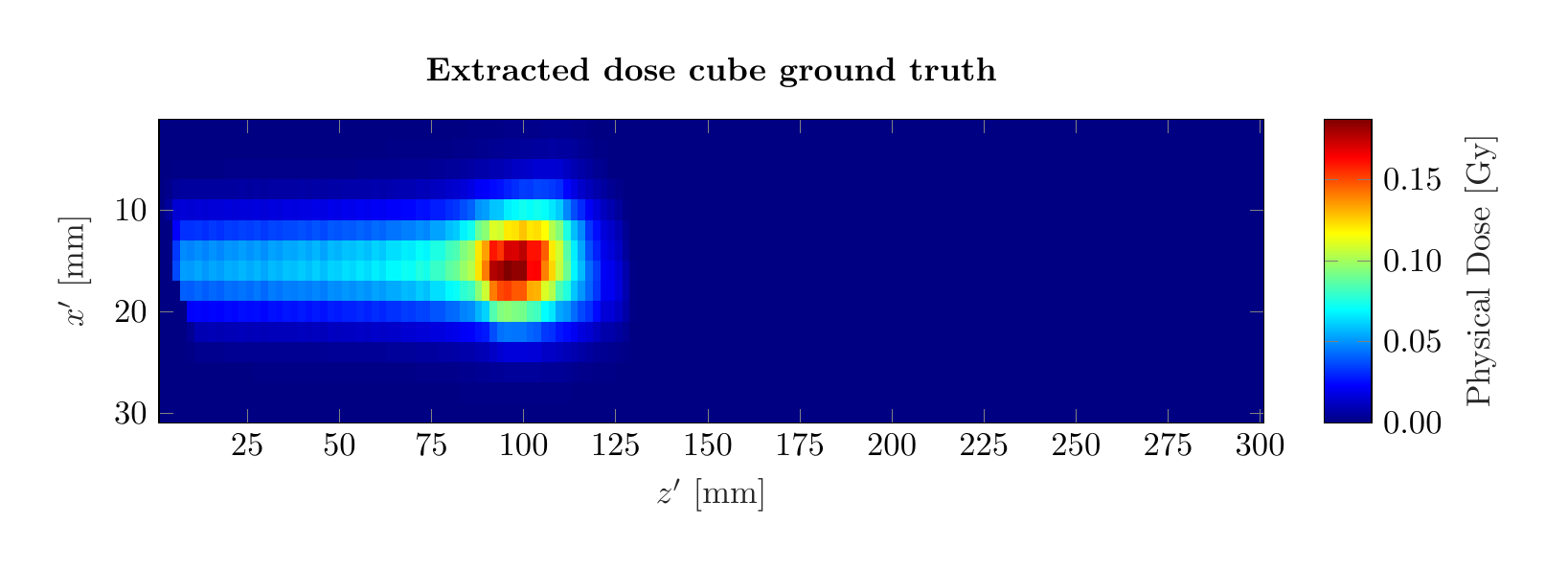}\label{fig:MandM_1c}} \\		
		\caption{(a) Dose distribution of a single pencil beam with initial energy 104.25 MeV impinging from gantry angle 240\textdegree\ overlaying the patient CT. The clipping region is highlighted with a red box. (b) Respective CT slice and (c) dose distribution in beam's eye view coordinate system.}
		\label{fig:cubeExtract}
	\end{center}
\end{figure}

The space of possible dose distributions can be further confined when switching from the patient coordinate system into the beam's eye view coordinate system. Here, the dose deposition is always oriented along the $z^{\prime}$-axis, as shown in figure \ref{fig:cubeExtract}. As the lateral extent and the particle range can be considered finite and is roughly known a priori for any given initial energy, it is further possible to perform a lateral and longitudinal clipping of the region of interest. In our case we use an isotropic resolution of \SI{2}{\milli\meter} with $m=15$ voxels in lateral direction and $l=150$ voxels in longitudinal direction (for patient setup).

Further, we chose to perform the learning not on HU maps but on maps of the relative stopping power (RSP)\footnote{The RSP denotes the range loss in the geometry relative to a water phantom.} which are also used for conventional pencil beam algorithms \cite{Wieser2017, Schaffner1999}. The RSP values are in turn translated into the respective water density for MC simulations.

In the described parameterization, we deal with a supervised regression problem that maps the geometry input data $\boldsymbol{\mathcal{G}}_i\in \mathbb{R}^{l\times m\times m}_{+}$ to real-valued dose outputs $\boldsymbol{\mathcal{D}}_{i}\in\mathbb{R}^{l\times m\times m}_{+}$.

\subsubsection{Network architecture considerations}

\begin{figure}[h]
	\centering
	\includegraphics[width=0.95\textwidth]{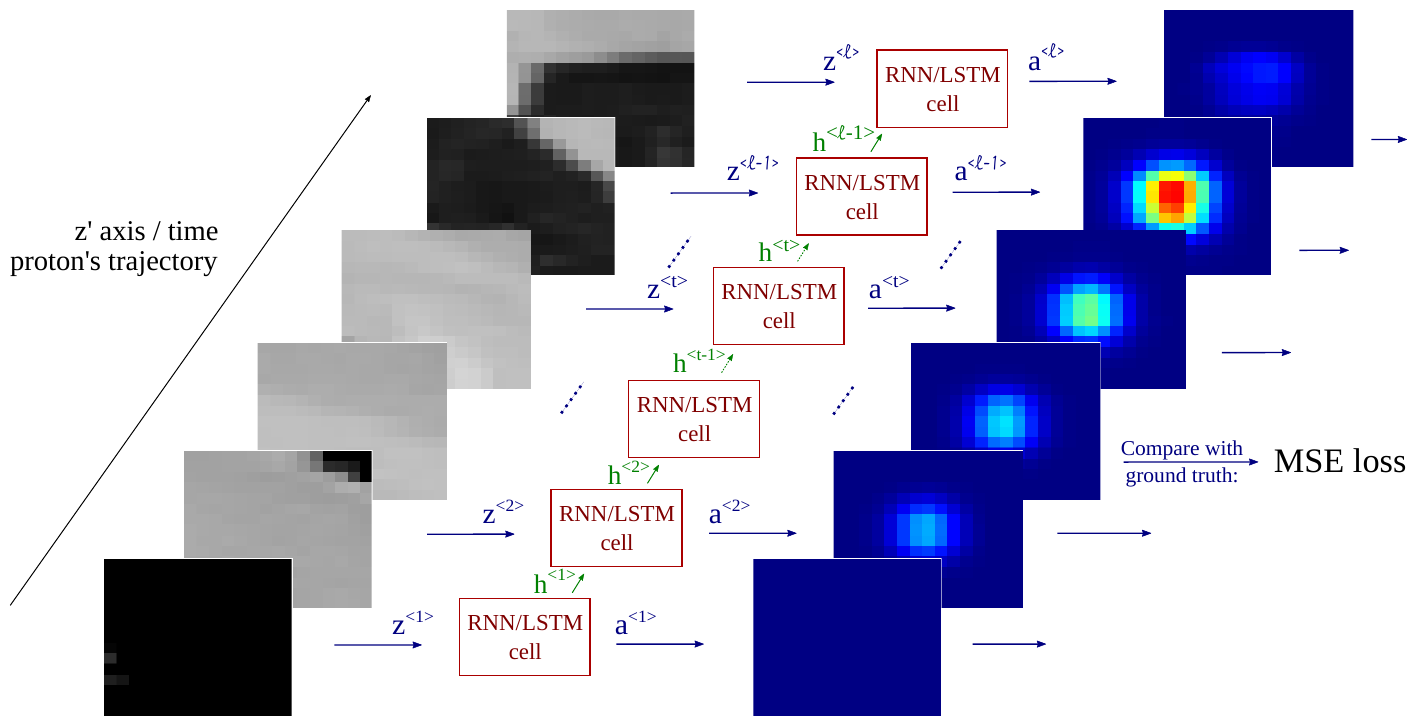}
	\caption{Sequential, spatio-temporal modeling of the proton dose calculation problem. each $m \times m$ slice of the input is flattened into a 1D input array $z^{<t>}$. Each input array is then passed to the RNN/LSTM \protect\footnotemark[1] cell generating a hidden inner state $h^{<t>}$ and an output $a^{<t>}$. The hidden inner state is passed as an input information for subsequent slices ($l$ slices in total), while the output is passed to a fully connected neural network back end to generate an $m \times m$ output slice. The output is then compared to the original ground truth by means of mean squared error loss.}
	\label{fig:3Dto2Dseq}
\end{figure}

Due to the \num{3}{D} nature of the problem, the intuitive implementation of a neural network for dose calculation is a \num{3}{D} model such as a \num{3}{D} U-net. These models have seen much attention recently with continuous advances in GPU hardware, and specifically GPU memory sizes, allowing processing of big data. However, the particle dose calculation problem exhibits a geometrical peculiarity that motivates a more specialized network architecture: dose deposition is almost exclusively taking place in a sequential \emph{upstream-to-downstream} manner. I.\,e. the highly energetic protons predominantly travel along one direction with moderate lateral scatter until they stop. This characteristic behavior allows for a representation of the \num{3}{D} input and output as a sequence of two-dimensional slices, as illustrated in figure \ref{fig:3Dto2Dseq}.

\begin{figure}[htb]
	\begin{center}
		\centering
		\subfigure[]{
%
%
\definecolor{mycolor1}{rgb}{0.00000,0.18750,1.00000}%
\definecolor{mycolor2}{rgb}{0.00000,0.75000,1.00000}%
\definecolor{mycolor3}{rgb}{0.37500,1.00000,0.62500}%
\definecolor{mycolor4}{rgb}{1.00000,1.00000,0.00000}%
\definecolor{mycolor5}{rgb}{1.00000,0.43750,0.00000}%
\definecolor{mycolor6}{rgb}{0.81250,0.00000,0.00000}%
\begin{tikzpicture}

\begin{axis}[%
width=4in,
height=1.064in,
at={(0in,2.08in)},
scale only axis,
point meta min=0,
point meta max=0.0653446093921575,
axis on top,
xmin=0.5,
xmax=60.5,
xtick={0.5,5.5,10.5,15.5,20.5,25.5,30.5,35.5,40.5,45.5,50.5,55.5,60.5},
xticklabels={{0},{},{20},{},{40},{},{60},{},{80},{},{100},{},{120}},
xlabel style={font=\color{white!15!black}},
xlabel={$\text{z}^\prime\text{ [mm]}$},
y dir=reverse,
ymin=0.5,
ymax=15.5,
ytick={0.5,3,5.5,8,10.5,13,15.5},
yticklabels={{0},{},{10},{},{20},{},{30}},
ylabel style={font=\color{white!15!black}},
ylabel={$\text{y}^\prime\text{ [mm]}$},
axis background/.style={fill=white},
title style={font=\bfseries, font=\small},
title={$\text{x}^\prime\text{ = 15 [mm]}$},
axis x line*=bottom,
axis y line*=left,
colormap/jet,
colorbar,
colorbar style={ylabel style={font=\color{white!15!black}, font=\tiny}, ylabel={Physical Dose [Gy]}}
]
\addplot [forget plot] graphics [xmin=0.5, xmax=60.5, ymin=0.5, ymax=15.5] {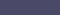};
\addplot [forget plot] graphics [xmin=0.5, xmax=60.5, ymin=0.5, ymax=15.5] {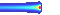};
\addplot [color=mycolor1, line width=1.0pt]
  table[row sep=crcr]{%
1	6.15612578919327\\
2	6.13834195879222\\
3	6.11666293578324\\
4	6.0971127105098\\
5	6.08287580214173\\
6	6.06640699441877\\
7	6.04616190893456\\
8	6.02880130514987\\
9	6.01185677186158\\
9.93939733811628	6\\
10	5.99916103813029\\
11	5.97669154193449\\
12	5.9558265050496\\
13	5.93578114224435\\
14	5.91951629734115\\
15	5.89940613644067\\
16	5.8776066795\\
17	5.85320745398254\\
18	5.83141420585365\\
19	5.8141048251827\\
20	5.78633573969092\\
21	5.7627460446625\\
22	5.73787620509039\\
23	5.70977509980655\\
24	5.67975380544962\\
25	5.64976411354567\\
26	5.61670323198598\\
27	5.58113726530283\\
28	5.54260054847562\\
29	5.50611730107015\\
30	5.45965298621925\\
31	5.41257193721095\\
32	5.35866528190333\\
33	5.29865216004894\\
34	5.23655923837093\\
35	5.16386508845441\\
36	5.08293952526568\\
36.7961238973755	5\\
37	4.96785766990063\\
38	4.76682756312474\\
39	4.506413586653\\
40	4.14911312884053\\
41	4.22111204076076\\
41.5134676316028	5\\
41.8009490321213	6\\
41.9018631587955	7\\
41.9286133249845	8\\
41.9030134830921	9\\
41.80234479086	10\\
41.5128306735451	11\\
41	11.7706718731885\\
40	11.8430697865932\\
39	11.4833445744237\\
38	11.2238361506457\\
37	11.0245565197191\\
36.8444704959333	11\\
36	10.9131290699096\\
35	10.8280429430365\\
34	10.7578200251545\\
33	10.6984071713918\\
32	10.6425968113925\\
31	10.5878266012864\\
30	10.5401563789385\\
29	10.499300620475\\
28	10.4579541835289\\
27	10.4209592727896\\
26	10.3851778577618\\
25	10.3505266287599\\
24	10.3201877786512\\
23	10.2896017733499\\
22	10.2609466358887\\
21	10.2342842879451\\
20	10.2111582966004\\
19	10.1831308974061\\
18	10.1644713738204\\
17	10.1419141550854\\
16	10.118253139978\\
15	10.0959664583084\\
14	10.0721273319105\\
13	10.053193786119\\
12	10.0338117705274\\
11	10.0144627085674\\
10.1255879466632	10\\
10	9.99814685116074\\
9	9.97932145605018\\
8	9.96410120954537\\
7	9.94919544160091\\
6	9.93481028427637\\
5	9.91302970448917\\
4	9.89303599044632\\
3	9.87772297297696\\
2	9.85960128793305\\
1	9.83873582853044\\
};

\addplot [color=mycolor2, line width=1.0pt]
  table[row sep=crcr]{%
41.49147976391	6\\
41.6934325354902	7\\
41.746728878987	8\\
41.6966609007561	9\\
41.493427605858	10\\
41	10.9188929593309\\
40	10.9914989586173\\
39	10.535779251477\\
38	10.2354036651118\\
37	10.0159857673937\\
36.9023818199677	10\\
36	9.84550072314148\\
35	9.70344587792953\\
34	9.58217635914066\\
33	9.48129674462171\\
32	9.38850646590638\\
31	9.3018786713033\\
30	9.22853493878102\\
29	9.15966659763556\\
28	9.10277912347254\\
27	9.04169007117925\\
26.1973942099934	9\\
26	8.97688719003396\\
25	8.87166447853481\\
24	8.7651582529804\\
23	8.66022074011281\\
22	8.5730080878356\\
21	8.4853326431615\\
20	8.3982364550372\\
19	8.31389171474482\\
18	8.23440011587935\\
17	8.16119846166727\\
16	8.07124940125701\\
15	8.00671828612336\\
14.9080967677349	8\\
15	7.99336895878033\\
16	7.92976816409114\\
17	7.84173227974008\\
18	7.7715778273629\\
19	7.69486521968143\\
20	7.61854288627396\\
21	7.52799971796096\\
22	7.43914850506611\\
23	7.35576811992078\\
24	7.24971329739778\\
25	7.15060397885755\\
26	7.04587023588483\\
26.4619447106357	7\\
27	6.97534098538175\\
28	6.91502579175299\\
29	6.85438577622853\\
30	6.78276312489029\\
31	6.70879954603563\\
32	6.62326399954836\\
33	6.52712776600512\\
34	6.42651931788865\\
35	6.30579913571271\\
36	6.16290397636157\\
36.9524843751126	6\\
37	5.99219736599938\\
38	5.76593956184126\\
39	5.46025620932315\\
40	5.00310917554027\\
41	5.07842271181789\\
41.49147976391	6\\
};

\addplot [color=mycolor3, line width=1.0pt]
  table[row sep=crcr]{%
41.1820104956988	6\\
41.4850019121849	7\\
41.5648444329895	8\\
41.4903083184201	9\\
41.184510420856	10\\
41	10.3436072984869\\
40	10.4363163558226\\
39.2788981926145	10\\
39	9.773782070947\\
38	9.35398274377432\\
37	9.04328704114542\\
36.8105193675333	9\\
36	8.55708646371865\\
35	8.12034307685822\\
34.6672297781211	8\\
35	7.88418469706635\\
36	7.4621442372565\\
36.8654671077444	7\\
37	6.96876498458189\\
38	6.65953091956841\\
39	6.22959162850043\\
39.2749410130674	6\\
40	5.55664897518677\\
41	5.65871079266344\\
41.1820104956988	6\\
};

\addplot [color=mycolor4, line width=1.0pt]
  table[row sep=crcr]{%
41.2765712888796	7\\
41.382959986992	8\\
41.2839557360842	9\\
41	9.7735995055986\\
40	9.88588038150819\\
39	9.02954334965203\\
38.9374494864042	9\\
38.1521985411606	8\\
38.9698936873861	7\\
39	6.98553801496289\\
40	6.10880881093304\\
41	6.2368679622459\\
41.2765712888796	7\\
};
\addplot [color=mycolor5, line width=1.0pt]
  table[row sep=crcr]{%
41.0681406655743	7\\
41.2010755409945	8\\
41.0776031537482	9\\
41	9.2114194352978\\
40	9.35286761768222\\
39.5919293482426	9\\
39.215327459205	8\\
39.608753953034	7\\
40	6.65541628065747\\
41	6.81198220110151\\
41.0681406655743	7\\
};

\addplot [color=mycolor6, line width=1.0pt]
  table[row sep=crcr]{%
41.019191094997	8\\
41	8.14464835140758\\
40	8.56245541044969\\
39.7623430797377	8\\
40	7.4596637378357\\
41	7.86448251041542\\
41.019191094997	8\\
};

\end{axis}
\end{tikzpicture}%
\label{fig:MandM_3a}} \\
		\subfigure[]{
%
%
\definecolor{mycolor1}{rgb}{0.00000,0.18750,1.00000}%
\definecolor{mycolor2}{rgb}{0.00000,0.75000,1.00000}%
\definecolor{mycolor3}{rgb}{0.37500,1.00000,0.62500}%
\definecolor{mycolor4}{rgb}{1.00000,1.00000,0.00000}%
\definecolor{mycolor5}{rgb}{1.00000,0.43750,0.00000}%
\definecolor{mycolor6}{rgb}{0.81250,0.00000,0.00000}%
\begin{tikzpicture}

\begin{axis}[%
width=4in,
height=1.064in,
at={(0in,0in)},
scale only axis,
point meta min=0,
point meta max=0.0653446093921575,
axis on top,
xmin=0.5,
xmax=60.5,
xtick={0.5,5.5,10.5,15.5,20.5,25.5,30.5,35.5,40.5,45.5,50.5,55.5,60.5},
xticklabels={{0},{},{20},{},{40},{},{60},{},{80},{},{100},{},{120}},
xlabel style={font=\color{white!15!black}},
xlabel={$\text{z}^\prime\text{ [mm]}$},
y dir=reverse,
ymin=0.5,
ymax=15.5,
ytick={0.5,3,5.5,8,10.5,13,15.5},
yticklabels={{0},{},{10},{},{20},{},{30}},
ylabel style={font=\color{white!15!black}},
ylabel={$\text{y}^\prime\text{ [mm]}$},
axis background/.style={fill=white},
title style={font=\bfseries, font=\small},
title={$\text{x}^\prime\text{ = 15 [mm]}$},
axis x line*=bottom,
axis y line*=left,
colormap/jet,
colorbar,
colorbar style={ylabel style={font=\color{white!15!black}, font=\tiny}, ylabel={Physical Dose [Gy]}}
]
\addplot [forget plot] graphics [xmin=0.5, xmax=60.5, ymin=0.5, ymax=15.5] {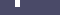};
\addplot [forget plot] graphics [xmin=0.5, xmax=60.5, ymin=0.5, ymax=15.5] {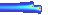};
\addplot [color=mycolor1, line width=1.0pt]
  table[row sep=crcr]{%
1	6.17978898765504\\
2	6.16151273721028\\
3	6.14050605710105\\
4	6.12102800708686\\
5	6.10742993732253\\
6	6.09251227041924\\
7	6.06972674201824\\
8	6.0511667925954\\
9	6.03561776818785\\
10	6.01925021240123\\
11	6.0012021523499\\
11.0606734801326	6\\
12	5.97978757049828\\
13	5.96110977301519\\
14	5.94196751896778\\
15	5.91880987759169\\
16	5.87254268424687\\
17	5.80368436373961\\
18	5.73521055487625\\
19	5.65691651341572\\
20	5.57067610505526\\
21	5.50859207998741\\
22	5.47422630051219\\
23	5.43351126445915\\
24	5.38918982431414\\
25	5.33618365675463\\
26	5.2791828294964\\
27	5.21371067798613\\
28	5.14049771477354\\
29	5.05923936169893\\
29.5450519009328	5\\
30	4.92402815756635\\
31	4.70572704461633\\
32	4.40369842006096\\
33	4.10285151087943\\
33.8887134329121	5\\
34	5.1787795211807\\
34.4341392308328	6\\
34.8453775384811	7\\
35	7.14716311333513\\
36	7.12703199817348\\
37	7.08284774169224\\
38	7.02056549063741\\
38.2146835149673	7\\
39	6.8433725148986\\
40	6.55104236174567\\
41	6.60920677386706\\
41.3897447414463	7\\
41.8671273952069	8\\
41.8923000174582	9\\
41.7958144732814	10\\
41.5011708711529	11\\
41	11.7543168802463\\
40	11.8234952860282\\
39	11.450985770177\\
38	11.1914387001019\\
37.0491437918218	11\\
37	10.9932227375336\\
36	10.8914274627583\\
35	10.8076096012239\\
34	10.7398051318758\\
33	10.6915085183564\\
32	10.6441186488353\\
31	10.5854504288841\\
30	10.5332130834261\\
29	10.4891212004219\\
28	10.4466315354995\\
27	10.4074679108308\\
26	10.3701095190467\\
25	10.3348016710928\\
24	10.3032443736038\\
23	10.2712801958372\\
22	10.242463852921\\
21	10.2177049171205\\
20	10.1891146762633\\
19	10.1624647238422\\
18	10.1435744402008\\
17	10.1242988839846\\
16	10.0981591512645\\
15	10.075190845026\\
14	10.0509766802519\\
13	10.0304718316749\\
12	10.010653478514\\
11.4757620323952	10\\
11	9.99141359457546\\
10	9.97873815597961\\
9	9.95895917625241\\
8	9.94353693762633\\
7	9.92758070860001\\
6	9.91402879407329\\
5	9.89190347093687\\
4	9.87251338405213\\
3	9.85639198048436\\
2	9.83805661527654\\
1	9.81524343538392\\
};

\addplot [color=mycolor2, line width=1.0pt]
  table[row sep=crcr]{%
33.1541350563708	5\\
33.8855525927809	6\\
34	6.36889069701917\\
34.2893047079566	7\\
35	7.67640969353563\\
36	7.64062906531343\\
37	7.57950023216419\\
38	7.49038226149254\\
39	7.34931539279647\\
40	7.10840117831441\\
41	7.15055989917096\\
41.6247320389838	8\\
41.6754662146767	9\\
41.4806045703891	10\\
41	10.8967004339049\\
40	10.9675468365563\\
39	10.4979026419349\\
38	10.1939286001492\\
37.105213576537	10\\
37	9.97513857802678\\
36	9.80441497242018\\
35	9.66557069389351\\
34	9.559786847725\\
33	9.50175246206253\\
32	9.40969627343155\\
31	9.30585977456037\\
30	9.22368365352541\\
29	9.14950328294311\\
28	9.08729147528413\\
27	9.02571610639201\\
26.5317561705049	9\\
26	8.94989305165353\\
25	8.85508094851129\\
24	8.76457709374991\\
23	8.65916359923436\\
22	8.57040212839886\\
21	8.47148473378913\\
20	8.37243603360096\\
19	8.27817783947641\\
18	8.17481717217117\\
17	8.07317880948105\\
16.1868648195894	8\\
17	7.90441592535314\\
18	7.70999656596924\\
19	7.32758288840511\\
19.3994813740509	7\\
20	6.91582532245013\\
21	6.82158032656563\\
22	6.77574793495452\\
23	6.72183365057111\\
24	6.65843774257259\\
25	6.5794874800906\\
26	6.48675486152818\\
27	6.37847657874745\\
28	6.24730463870521\\
29	6.07917533710001\\
29.3295316520669	6\\
30	5.87542868842235\\
31	5.63648258471341\\
32	5.28236850393651\\
32.6744520201822	5\\
33	4.84440199976442\\
33.1541350563708	5\\
};

\addplot [color=mycolor3, line width=1.0pt]
  table[row sep=crcr]{%
33.5092051128069	6\\
33.8389415308796	7\\
34	7.93449800444468\\
34.079643542805	8\\
34	8.06230314985224\\
33	8.57921826258752\\
32	8.28285407457301\\
31.1494214730983	8\\
31	7.88216454951622\\
30.0244028601645	7\\
31	6.61985822014516\\
31.8410081991453	6\\
32	5.93192497267474\\
33	5.36868959510975\\
33.5092051128069	6\\
};

\addplot [color=mycolor3, line width=1.0pt]
  table[row sep=crcr]{%
41.1653946674968	10\\
41.4586324118951	9\\
41.3823366827608	8\\
41	7.48014174063611\\
40	7.4188596552169\\
39	7.76522847419506\\
38	7.96019903234768\\
37.6441223750227	8\\
37.1010947180513	9\\
38	9.27249659966706\\
39	9.705186291766\\
39.345117168326	10\\
40	10.3980022846859\\
41	10.3085893877162\\
41.1653946674968	10\\
};

\addplot [color=mycolor4, line width=1.0pt]
  table[row sep=crcr]{%
33.1328576328329	6\\
33.5032183644771	7\\
33	7.86207862776522\\
32	7.09373302877724\\
31.8959018106648	7\\
32	6.92109119670163\\
32.7171815862341	6\\
33	5.83528365315467\\
33.1328576328329	6\\
};

\addplot [color=mycolor4, line width=1.0pt]
  table[row sep=crcr]{%
41.2417986091135	9\\
41.1399413265378	8\\
41	7.80972358210125\\
40	7.72931813211938\\
39.3331235069064	8\\
39.0785263518248	9\\
40	9.82272486930517\\
41	9.7011562750624\\
41.2417986091135	9\\
};

\addplot [color=mycolor5, line width=1.0pt]
  table[row sep=crcr]{%
33.1674951980746	7\\
33	7.28694109894704\\
32.6492038335878	7\\
33	6.56461038414605\\
33.1674951980746	7\\
};

\addplot [color=mycolor5, line width=1.0pt]
  table[row sep=crcr]{%
41.024964806332	9\\
41	8.78592049960145\\
40	8.24360651430162\\
39.7377498281814	9\\
40	9.23414640099799\\
41	9.07239177545125\\
41.024964806332	9\\
};

\end{axis}
\end{tikzpicture}%
\label{fig:MandM_3b}}	
		\caption{Effect of a heterogeneity on the shape of a pencil beam dose profile. The pencil beam is formed by \num{e6} protons with an initial energy of \SI{104.25}{\mega\electronvolt} passing through (a) water (\SI{1.0}{RSP}). (b) The cuboid heterogeneity (\SI{2.5}{RSP}) has \SI{10}{\milli\meter} width in $z^{\prime}$ axis, \SI{14}{\milli\meter} width in $x^{\prime}$ axis, and \SI{1}{\milli\meter} distance to the center of the of proton beam. The effect of the cuboid mainly manifests in a bimodal Bragg peak region extending from $\approx$ \SIrange{60}{80}{\milli\meter}, i.\,e., $20mm$ after the heterogeneity.}
		\label{fig:informationPass}
	\end{center}
\end{figure}

Thence, the dose calculation problem has strong similarities to conventional video analysis in terms of spatio-temporal features. In action recognition tasks for instance, models have to extract spatial features of objects within each frame, and temporal features to interpret the movement of those objects. Simulating the protons traverse through matter, and consequently their dose deposition, is very similar to this task. It is completely determined by the upstream geometry, i.\,e.,  the geometry previously "seen" by the protons along their track through the patient. This implies causality from upstream to downstream within the \num{3}{D} volume and it suggest a special role for regions in the input data that have high gradients in their RSP values (e.\,g. material interfaces to bones with high RSP and cavities with low RSP). Thereby, the effect of each heterogeneity on the dose deposition is most pronounced at the end of the proton range as demonstrated in figure \ref{fig:informationPass}. Consequently, any model to simulate dose deposition for particles needs to extract spatio-temporal features and precisely propagate the impact of heterogeneities along the particle tracks.

While recent studies \cite{Kim2019, Hou, Ye} have shown that \num{3}{D-CNN} models are capable of extracting spatio-temporal features through sequential data, consideration has to be taken with regard to the length of the sequences. In our application domain protons may have high ranges of more than \SI{300}{\milli\meter} and consequently very high number of slices which may lead to issues with an adequate receptive field for detecting such long coherence. The \num{3}{D-CNN} by design has the disadvantage of having high computational complexity and excessive memory usage \cite{Kim2019}. In order to capture temporal coherence, the total number of parameters of \num{3}{D-CNN}s will increase manifolds depending on the size of the temporal receptive field.

Processing sequences with long dependencies requires a model capable of passing information through the series. Recurrent Neural Networks (RNNs) with their hidden inner states, are capable of connecting many conventional one-input-to-one-output neural networks resulting in a model suitable to process on a many-input-to-many-output layouts. LSTM networks, an evolved version of simple RNNs, are capable of effectively transmitting relevant information through very long series thanks to their internal mechanism. Moreover, one directional LSTM models can fully adapt to the upstream-to-downstream propagation scheme of protons, eliminating dependencies between downstream to upstream resulting in a substantially reduced number of parameters for the model.

\footnotetext[1]{Note that this figure is illustrating the RNN/LSTM network in an \emph{unfolded} fashion \cite{LeCun2015}.}

\subsubsection{LSTM networks}
\label{sec:LSTM networks}

RNN models are distinguished by having a hidden internal state, enabling them to retain temporal information, similar to a memory. Considering our problem parameterization, depicted in figure \ref{fig:3Dto2Dseq} of having an input sequence $\left\{\boldsymbol{z}_{1}, \ldots, \boldsymbol{z}_{t-1}, \boldsymbol{z}_{t}, \boldsymbol{z}_{t+1}, \ldots, \boldsymbol{z}_{l}\right\}$ and output sequence $\left\{\boldsymbol{a}_{1}, \ldots, \boldsymbol{a}_{t-1}, \boldsymbol{a}_{t}, \boldsymbol{a}_{t+1}, \ldots, \boldsymbol{a}_{l}\right\}$ with $\boldsymbol{z}_{t}=\left[z_{1}, \ldots, z_{m^2}\right]$ and $\boldsymbol{a}_{t}=\left[a_{1}, \ldots, a_{n}\right]$, RNNs calculate a hidden inner state $\boldsymbol{h}_{t} \in \mathbb{R}^{K}$ with K hidden units, and the output $\boldsymbol{a_t}$ via the following recurrence equations:

\begin{equation}\begin{array}{l}
\boldsymbol{h}_{t}=f\left(\boldsymbol{W}_{I H} \boldsymbol{z}_{t}+\boldsymbol{W}_{H H} \boldsymbol{h}_{t-1}+\boldsymbol{b}_{H}\right), \\
\boldsymbol{a}_{t}=f\left(\boldsymbol{W}_{H O} \boldsymbol{h}_{t}+\boldsymbol{b}_{O}\right),
\end{array}\end{equation}

where f denotes an element-wise non-linear activation function. $\boldsymbol{W}_{IH}$, $\boldsymbol{W}_{HH}$ and $\boldsymbol{W}_{HO}$ are the input-to-hidden, hidden-to-hidden, and hidden-to-output weight matrices, respectively. $\boldsymbol{b}_H$ and $\boldsymbol{b}_O$ denote the hidden and output layer biases. The weight matrices and the biases are shared parameters that, given a properly converging training process during backpropagation, map relations between the input and output as desired. Simple RNNs, however, often suffer from the vanishing or exploding gradient problem during backpropagation with an increasing number of events in the input sequence. These problems arise due to the recursive derivative operations taken place along the sequence which may lead to very small or very big gradients, which in turn disrupt the training process and consequently restrict the RNN's capabilities to memorize long-term dependencies. 

\begin{figure}[htb]
	\begin{center}
		\centering
		\includegraphics[width=0.75\linewidth]{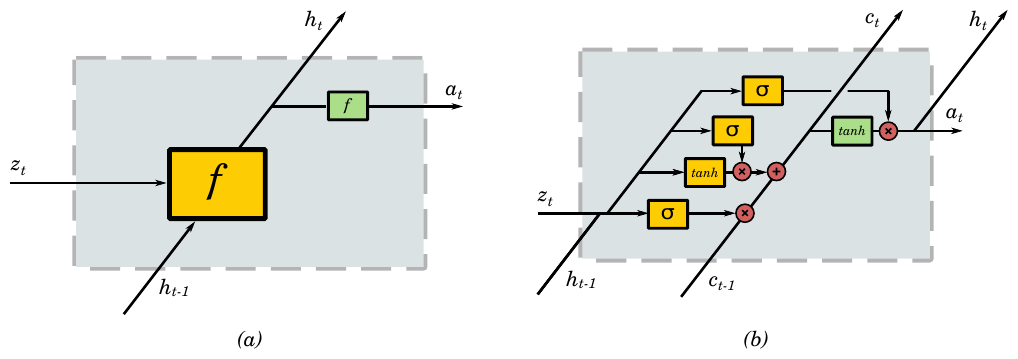}
		\label{fig:MandM_4}
		\caption{A schematic diagram of the internal module for (a) simple RNN cell and (b) for an LSTM cell.}
		\label{fig:LSTM}
	\end{center}
\end{figure}

With the motivation to overcome the long-term dependency problem, LSTM networks were first introduced by Hochreiter et al. \cite{Hochreiter1997}. The key innovation in this context is a memory cell state $\boldsymbol{c}_t$ and associated update mechanisms via gates in addition to the hidden state $\boldsymbol{h}_t$. In particular, the memory cell state has the capability of remaining unaltered, unless the three individually trained neural network layers, namely the \emph{input}, \emph{forget}, and \emph{cell} gate determine to update the information within memory cell state. This mechanism inhibits repetitive multiplication of the gradients in the course of backpropagation algorithm and ensures efficient passage for relevant information through the sequence, as shown in figure \ref{fig:LSTM}. Finally, an additional neural network, the \emph{output gate}, is trained to select the corresponding information to output for the current time interval. Mathematically, each LSTM cell enclosing the mentioned gates can be described via the following equations

\begin{equation}\begin{aligned}
\boldsymbol{i}_{t} &=\sigma\left(\boldsymbol{W}_{i_1} \boldsymbol{z}_{t}+\boldsymbol{W}_{i_2} \boldsymbol{h}_{t-1}+\boldsymbol{b}_{i}\right) \\
\boldsymbol{f}_{t} &=\sigma\left(\boldsymbol{W}_{f_1} \boldsymbol{z}_{t}+\boldsymbol{W}_{f_2} \boldsymbol{h}_{t-1}+\boldsymbol{b}_{f}\right) \\
\boldsymbol{g}_{t} &=\tanh \left(\boldsymbol{W}_{g_l} \boldsymbol{z}_{t}+\boldsymbol{W}_{g_2} \boldsymbol{h}_{t-1}+\boldsymbol{b}_{g}\right) \\
\boldsymbol{o}_{t} &=\sigma\left(\boldsymbol{W}_{o_1} \boldsymbol{z}_{t}+\boldsymbol{W}_{o_2} \boldsymbol{h}_{t-1}+\boldsymbol{b}_{o}\right) \\
\boldsymbol{c}_{t} &=\boldsymbol{f}_{t} \odot \boldsymbol{c}_{t-1}+\boldsymbol{i}_{t} \odot \boldsymbol{g}_{t} \\
\boldsymbol{h}_{t} &=\boldsymbol{o}_{t} \odot \tanh \left(\boldsymbol{c}_{t}\right),
\end{aligned}\end{equation}

where $\boldsymbol{W}_{\xi_1}$, $\boldsymbol{W}_{\xi_2}$, $\boldsymbol{b}_\xi$, $\xi \in \{i, f, g, o\}$, are the input-to-hidden weight matrices, hidden-to-hidden weight matrices, and biases that jointly are the learnable parameters constituting the input, forget, cell, and output gates, respectively. $\sigma$ is the Sigmoid function restricting the outputs to values between zero and one, ensuring a functionality analogous to gates. $\odot$ denotes the element-wise Hadamart product. The gates regulate the error propagation in the training process, thereby preventing the vanishing and exploding of the derivatives. Many variants of the original LSTM \cite{Graves2005, Gers, Gers2000RecurrentNT} have been introduced so far, and this study is using the Pytorch\footnote[1]{https://pytorch.org/docs/stable/nn.html\#lstm} implementation of this architecture.

\subsubsection{LSTM Training}
Training of the network was carried out with an Adam optimizer \cite{kingma2014adam}, with a learning rate of \num{e-5} and a Mean Squared Error (MSE) loss function. The LSTM cells featured one layer with \num{1000} neurons as internal layer, followed by a fully connected neural network for the back end. The back end network features one hidden layer with \num{100} neurons and an output layer of $m^2$ to generate the slices. The dose cubes were normalized to have values in \numrange{0}{1} range. Empirically, we found no improvement in test loss after about \num{100} epochs, and after that overfitting of the training set has been observed. Training of the network takes \numrange{3}{4} hours for the patient data set described in the next section, with a Geforce GTX \num{970} GPU.

\subsection{Dosimetric evaluation}
\label{sec:dosimetricevaluatoni}

\subsubsection{Phantom cases}

\begin{figure}[h]
	\centering
	\includegraphics[width=0.8\textwidth]{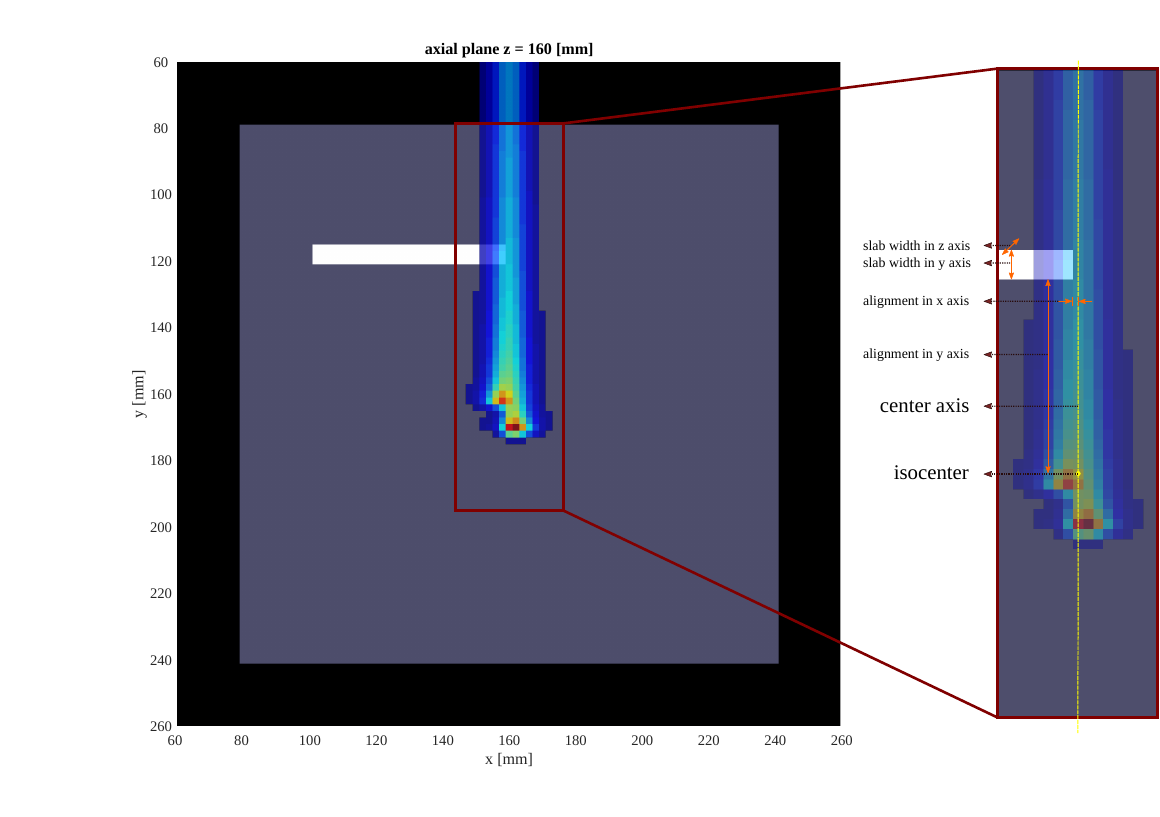}
	\caption{Phantom case setup; different geometric problems were generated by varying the slabs' dimensions in y and z axis, varying the alignment of the slab in x and y axis, and varying the density of both the water and the slab.}
	\label{fig:waterbox}
\end{figure}

In order to study the performance of the proposed neural network dose calculation algorithms in an idealized setting, we first carried out simulations on phantom geometries featuring cuboid inhomogeneities of varying dimensions and densities placed randomly within a water phantom, as shown in figure \ref{fig:waterbox}. For this task, \num{2500} phantom samples were generated with corresponding dose distributions from TOPAS Monte Carlo simulations. This number was raised to \num{10000} samples by augmenting rotated (\ang{90} angles) replicas of the cubes. \num{8000} samples were used as training set, and \num{2000} samples were used as test set. All the samples were simulated with $\sim\num{1.1e6}$ histories on average, resulting in less than \SI{1}{\percent} statistical uncertainty\footnote[2]{Uncertainty is calculated by dividing the highest standard deviation by the dose in that voxel}. 

\subsubsection{Lung cases}

\begin{figure}[htb]
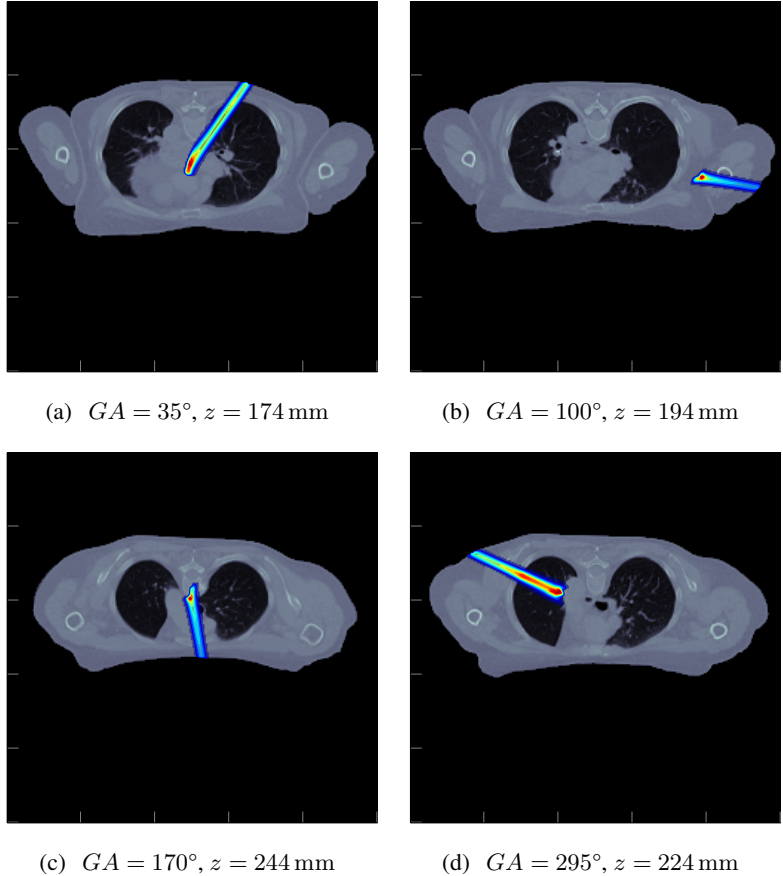

	\begin{center}
		\centering
		\subfigure[\hspace{1mm}$GA = $ \ang{35}, $z = $ \SI{174}{\milli\meter}]{\input{./imgs/lungDose_a.tikz}\label{fig:lungDose_a}} 
		\subfigure[\hspace{1mm}$GA = $ \ang{100}, $z = $ \SI{194}{\milli\meter}]{\input{./imgs/lungDose_b.tikz}\label{fig:lungDose_b}}\\
		\subfigure[\hspace{1mm}$GA = $ \ang{170}, $z = $ \SI{244}{\milli\meter}]{\input{./imgs/lungDose_c.tikz}\label{fig:lungDose_c}} 
		\subfigure[\hspace{1mm}$GA = $ \ang{295}, $z = $ \SI{224}{\milli\meter}]{\input{./imgs/lungDose_d.tikz}\label{fig:lungDose_d}}		
		\caption{Lung case setup; generating different geometric problems for preparing training data set by varying the gantry angles ($GA$) and shifting the isocenter along the $z$ axis.}
		\label{fig:patientSetup}
	\end{center}
\end{figure}

In order to study the performance of the proposed neural network dose calculation algorithms for real-world patient cases, we further considered dose calculation tasks on lung patient cases exhibiting highly pronounced inhomogeneities between normal tissue, lung tissue, and bony anatomy (rib cage \& spine). For this task, \num{1000} lung case samples were generated with corresponding dose distributions from TOPAS Monte Carlo simulations (\num{4000} after data augmentation). All samples stem from the same patient. Different geometric problems could be extracted from one patient by sampling the beam orientation in \ang{5} steps from \ang{0} to \ang{355} in combination with isocenter position samples in \SI{10}{\milli\meter} shifts spanning the lung along the $z$ axis, as shown in Fig. \ref{fig:patientSetup}. All the samples were simulated with $\num{2.5e6}$ histories on average, ensuring a statistical uncertainty between \SIrange{1}{2}{\percent}. \num{3200} samples were used as training set, 800 samples were used as test set. The original CT was downsampled to an isotropic \SI{2}{\milli\meter} resolution. Consequently, the resulting HU map was transformed to an \SI{}{RSP} map via \si{HU} look up tables yielding \si{RSP} values between \num{0} for vacuum and \num{2.5} for denser bone structures. 

\subsubsection{$\gamma$-index analysis}

In order to compare \num{3}{D} dose distributions, $\gamma$-analysis \cite{Low1998} was performed with a \SI{0.5}{\percent} distance-to-agreement and \SI{1}{\milli\meter} dose difference criterion([\SI{0.5}{\percent} , \SI{1}{mm}] in short) for the water box phantom, and a [\SI{0.5}{\percent} , \SI{2}{mm}] criterion was chosen for the patient case. The $\gamma$-analysis represents the agreement of the two \num{3}{D} dose distributions with a \num{3}{D} numerical index $\gamma$, having values less than \num{1} for voxels which \emph{pass} the agreement requirements, and values higher than \num{1} for voxels which \emph{fail} the agreement requirements. Consequently, the $\gamma$-index pass rate demonstrates the percentage of voxels with $\gamma < 1$ out of all voxels with $\gamma$-index value higher than zero (we do not consider all voxels to avoid inclusion of passing voxels beyond the range of the dose).

Unlike of dose difference maps, the $\gamma$-index provides a more holistic assessment of dose distribution differences, not only considering local dosimetric discrepancies but also spatial shifts (e.\,g. due to range offsets). Although computationally expensive, the $\gamma$-index pass rate reduces the discrepancy of two \num{3}{D} dose distributions to a single number which facilitates the large scale comparisons needed for our study working with several thousand training and test samples.

\section{Results}
\label{sec_results}

\subsection{Phantom cases}

\begin{figure}[h]
	\centering
	\includegraphics[width=0.65\textwidth]{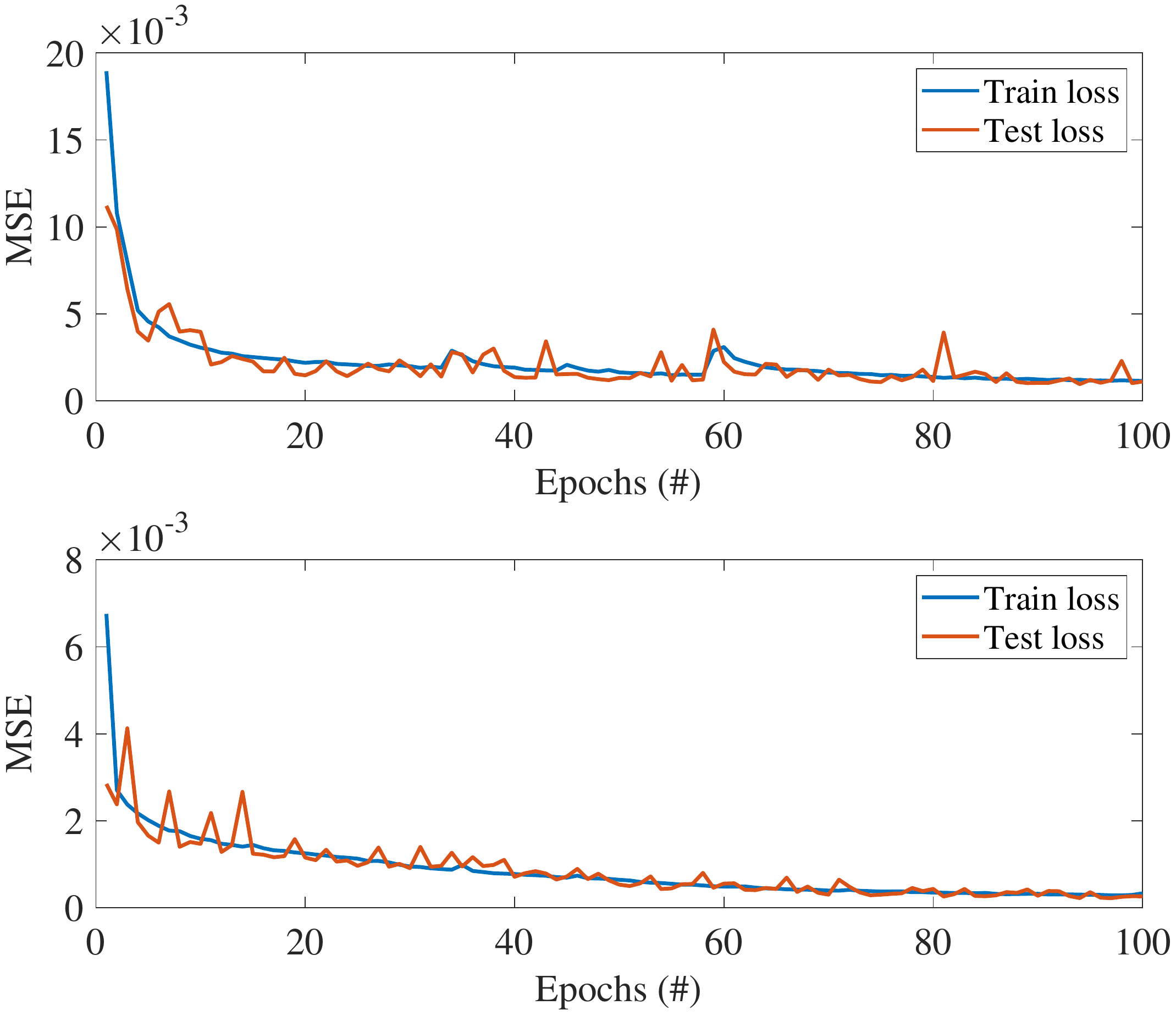}
	\caption{MSE loss function for training of the water box phantom data set for both the simple RNN architecture (top), and the LSTM architecture (bottom)}
	\label{fig:lossPlotBox}
\end{figure}

The prepared dataset for the water box phantom was used for training of the simple RNN and the LSTM network. Figure \ref{fig:lossPlotBox} shows MSE loss plot for the training of both architectures for \num{100} epochs. 

\begin{figure}[h]
	\centering
	\includegraphics[width=0.65\textwidth]{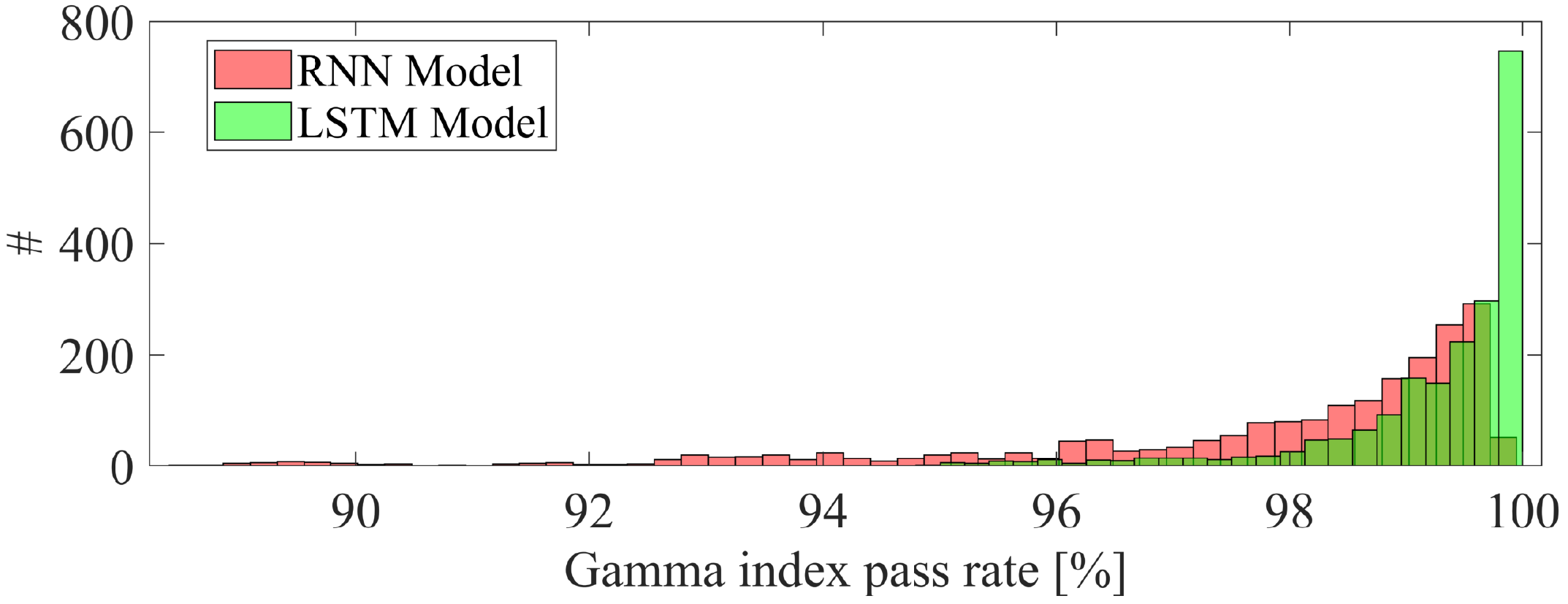}
	\label{fig:histogramCompare}
	\caption{Comparison between the RNN and LSTM model $\gamma$-index pass rate distribution over all test cases.}
\end{figure}

\begin{figure}[htb]
	\begin{center}
		\centering
		\subfigure[\scriptsize{ Input CT (top), ground truth MC calculation (bottom)}]{\includegraphics[width=0.49\linewidth]{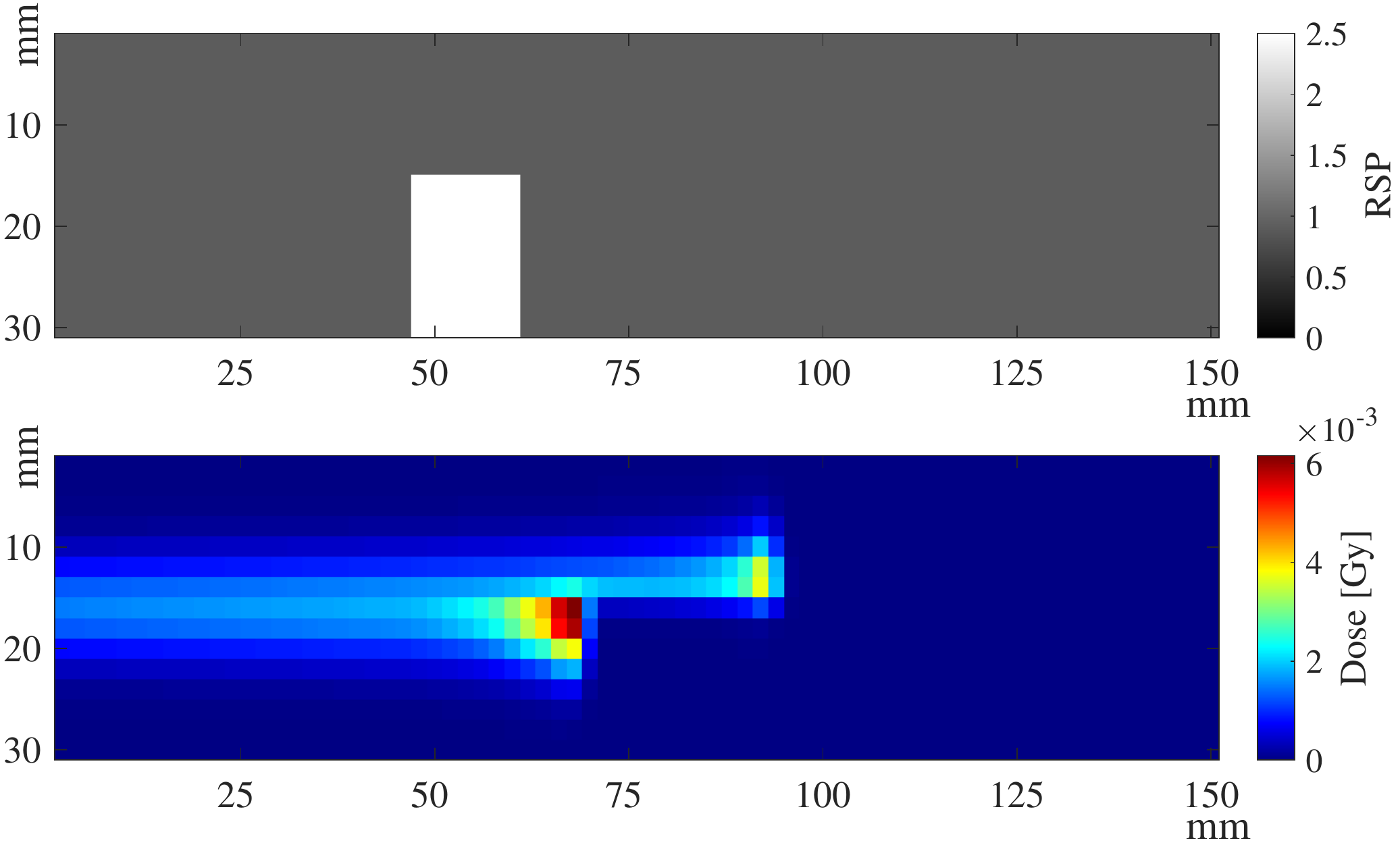}\label{fig:gammaMap2by2_MC}} \\
		\subfigure[\scriptsize{ RNN dose estimation (top), $\gamma$ map (bottom) \newline $\gamma$-index pass rate = 89.4 \%}]{\includegraphics[width=0.49\linewidth]{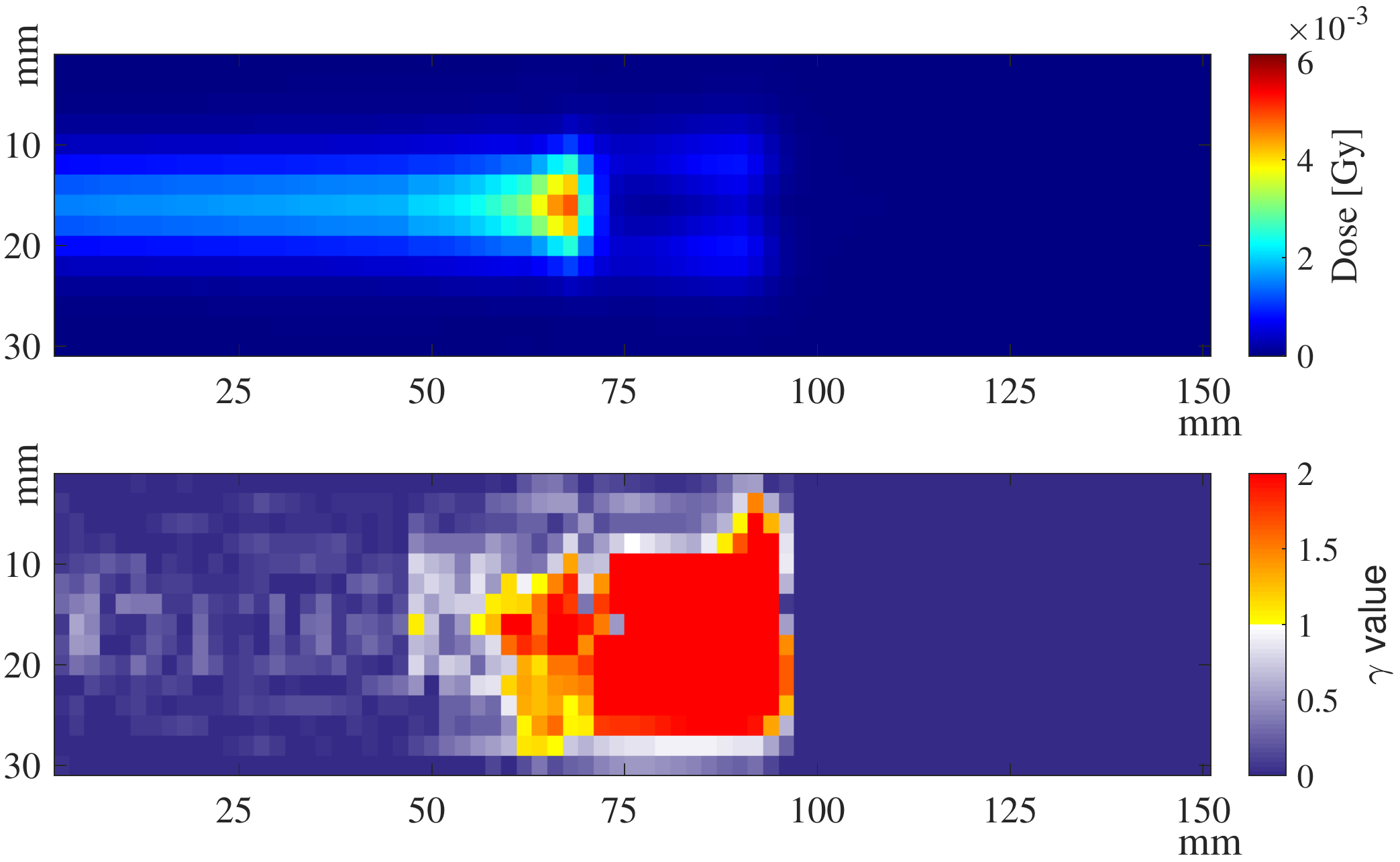}\label{fig:gammaMap2by2_RNN}}
		\hfill
		\subfigure[\scriptsize{ LSTM dose estmiation (top), $\gamma$ map (bottom) \newline $\gamma$-index pass rate = 97.1 \%}]{\includegraphics[width=0.49\linewidth]{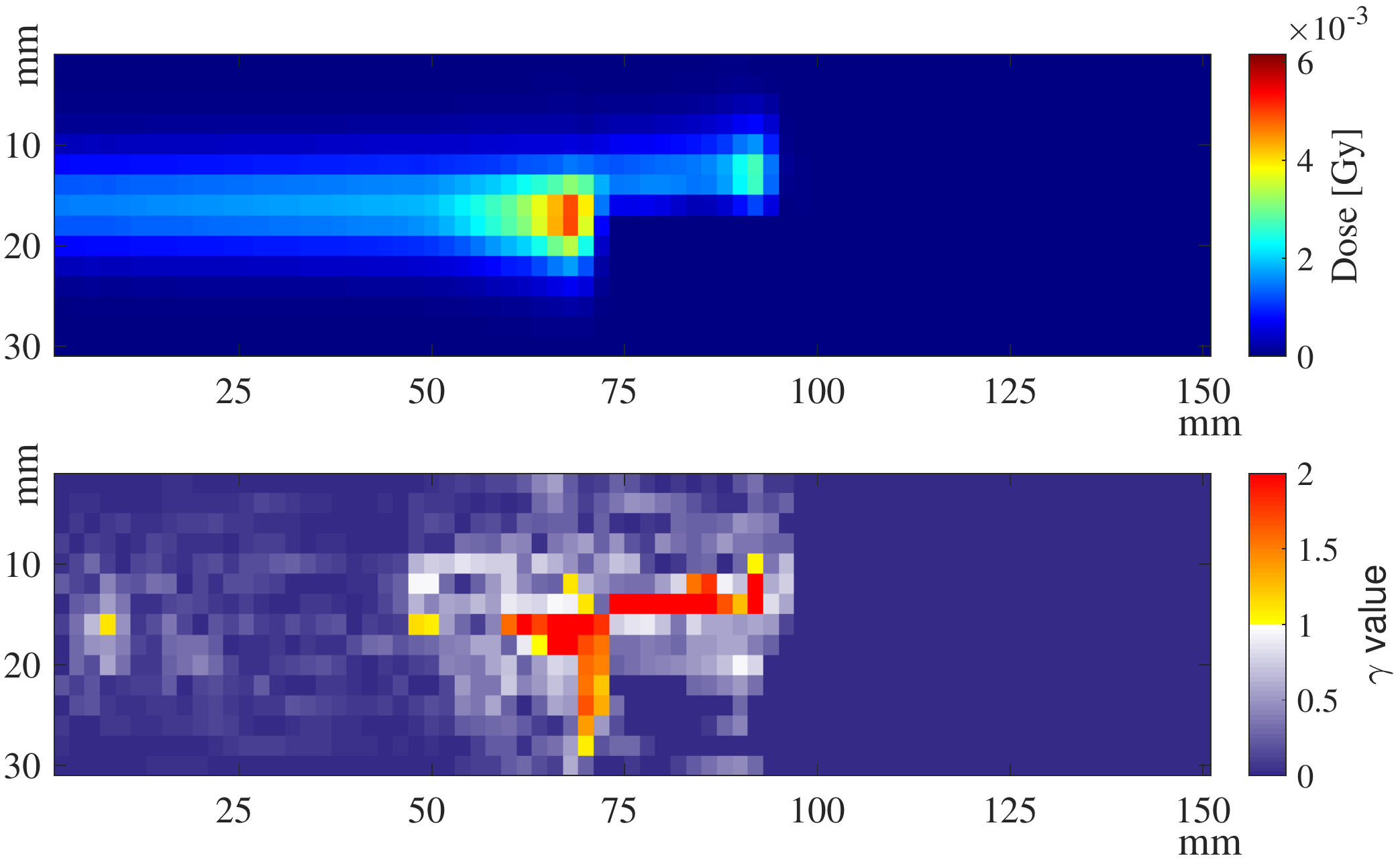}\label{fig:gammaMap2by2_LSTM}}
		\caption{\label{fig:RNNvsLSTMsimpleExmaple}Performance comparison of the (b) RNN and the (c) LSTM network with (a) ground truth MC calculation for a sample test data (\SI{104.25}{\mega\electronvolt}, with a \SI{14}{\milli\meter} width slab and \SI{2.5}{RSP}, $\gamma$-analysis criterion = [\SI{0.5}{\percent} , \SI{1}{mm}]) }
	\end{center}
\end{figure}

The performance of the two networks was further evaluated dosimetrically for the test set. Table \ref{tbl:statisticsOfArhcitectures} presents the outcome of the $\gamma$-analysis comparing the estimated dose from the networks with the ground truth MC calculations. While both networks seem generally suited for dose calculation with mean pass rates $>$\SI{97.88}{\percent} (figure \ref{fig:histogramCompare}), the LSTM network outperforms the RNN by \SI{1.5}{\percent}. We have observed that differences between the LSTM network and RNN mainly originate from cases with pronounced heterogeneities as shown in Figure \ref{fig:RNNvsLSTMsimpleExmaple}. In this example, the LSTM model demonstrates an evident improvement in comparison to the RNN model, which fails to predict the bimodal Bragg peak behind the density interface resulting in an $\sim8$ percentage points increase in overall $\gamma$-index pass rate.

\begin{table}[h]
	
	\begin{minipage}{0.999\textwidth}
		\centering	
		\caption{\label{tbl:statisticsOfArhcitectures}
			$\gamma$-index analysis comparing the two trained models in water phantom case ([0.5\%, 1mm]).}		
		\begingroup
		\setlength{\tabcolsep}{14pt}
		\renewcommand{\arraystretch}{1.5}
		\begin{tabular}{lcccc}
			\hline\hline
			&mean&std&min&max \\ [0.5ex]
			\hline
			RNN& 97.88 & 2.12 & 89.42 & 99.8\\ 
			LSTM& 99.29 & 0.8834 & 94.8 & 100\\ [1ex]
			\hline\hline
		\end{tabular}
		\endgroup
	\end{minipage}
\end{table}

\subsection{Patient cases}

\begin{figure}[h]
	\centering
	\includegraphics[width=0.65\textwidth]{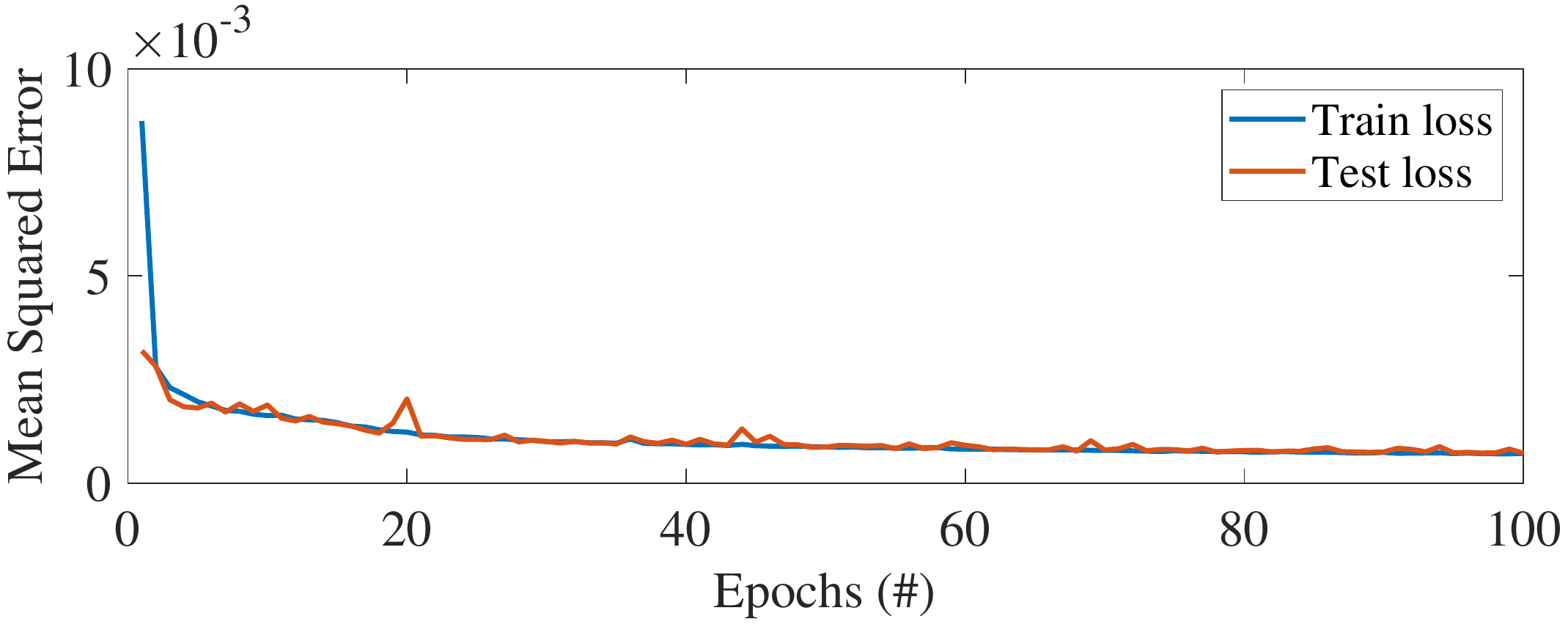}
	\caption{MSE loss function for training of the patient case data set with LSTM architecture}
	\label{fig:lossLSTM_patient}
\end{figure}

Due to the promising results of the LSTM network on the phantom cases, especially those with pronounced heterogeneities, we also trained this architecture on the lung patient data set, as shown in Figure \ref{fig:lossLSTM_patient}. Table \ref{tbl:gammaIndexPassRatePatient} summarizes the outcome of the $\gamma$-analysis for the set-aside test set. Note that we also included a relaxed $\gamma$ criterion of [\SI{0.5}{\percent} , \SI{2}{mm}] for the patient case. Due to the generation of the training and test set samples for the patient case by varying, among others, gantry angles, we have to deal with interpolation effects in the cube extraction that effect the $\gamma$-analysis at the previously used [\SI{0.5}{\percent} , \SI{1}{mm}] criterion.

\begin{table}[h]	
	\begin{minipage}{0.99\textwidth}
		\centering
		\caption{\label{tbl:gammaIndexPassRatePatient}
			$\gamma$-index analysis comparing the trained LSTM model with MC calculations for the patient case}
		\begingroup
		\setlength{\tabcolsep}{14pt}
		\renewcommand{\arraystretch}{1.5}
		\begin{tabular}{lcccc}
			\hline\hline
			$\gamma$-analysis criteria&mean&std&min&max \\ [0.5ex]
			\hline
			$[\SI{0.5}{\percent} , \SI{1}{mm}]$& 94.47 & 3.78 & 80.43 & 99.58\\ 
			$[\SI{0.5}{\percent} , \SI{2}{mm}]$& 99.33 & 0.92 & 94.91 & 100\\
			\hline\hline
		\end{tabular}
		\endgroup
	\end{minipage}
\end{table}

\begin{figure}[h]
	\centering
	\includegraphics[width=0.6\textwidth]{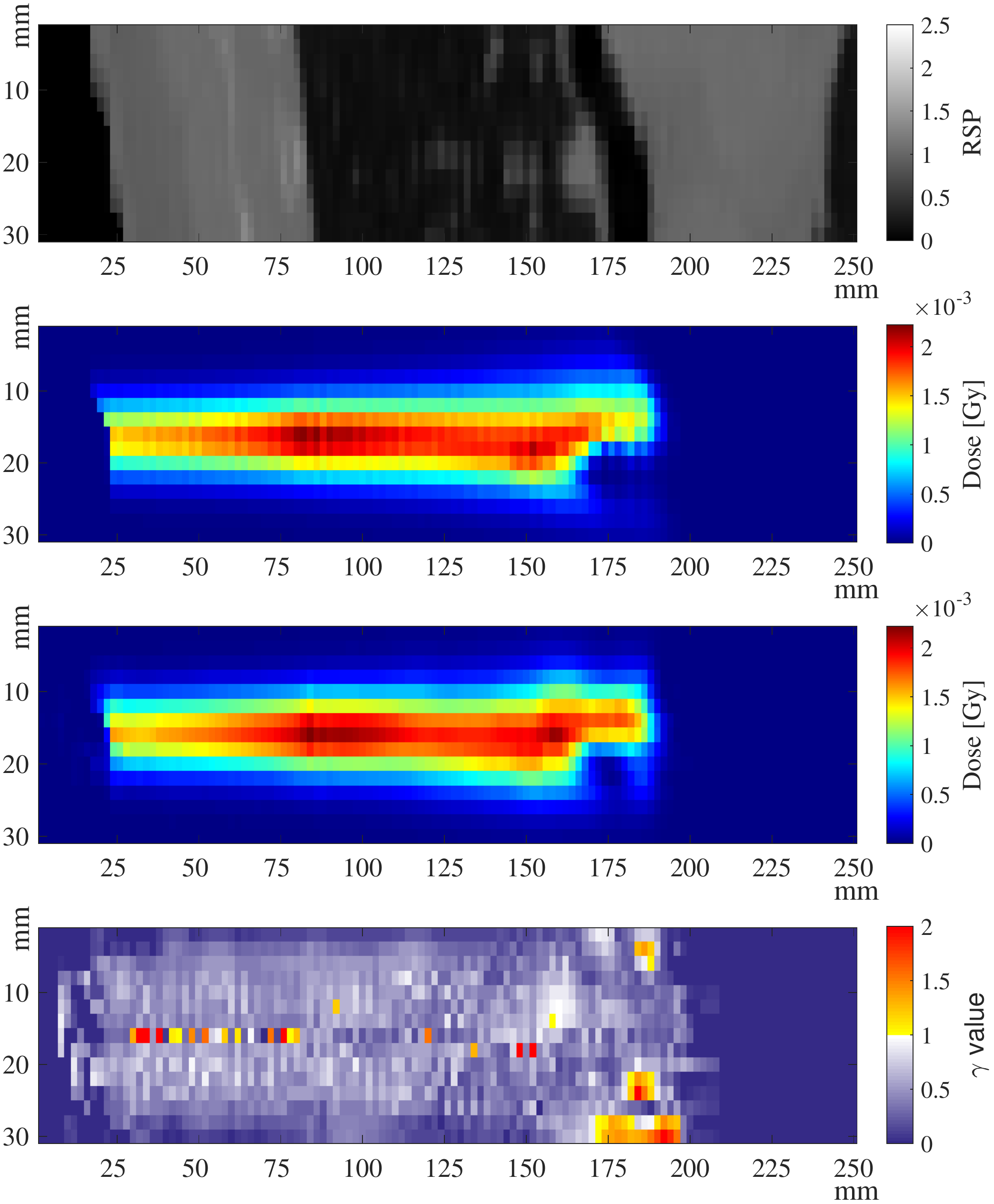}
	\caption{Dose estimation result for a sample test data (\SI{104.25}{\mega\electronvolt}). Starting from top is the input patient CT, the ground truth MC dose distribution, the estimated dose by the LSTM network, and the $\gamma$-index map ([\SI{0.5}{\percent} , \SI{2}{mm}])}
	\label{fig:LSTMpatientShowcase}
\end{figure}

Figure \ref{fig:LSTMpatientShowcase} shows the performance of the trained network on a representative test sample. In particular, we want to point out the capability of the trained network to deal with oblique gantry angles where voxels with vanishing density prior to entering the patient are successfully recognized and not confused with low density lung voxels lying within the patient. Furthermore, the LSTM network correctly predicts a smeared out Bragg peak without a distinct maximum at the end of the particles' range which is due to low density lung tissue at the location of the Bragg peak. Also the irregular shape of the distal fall-off which originates from inhomogeneities in the pencil beam track is qualitatively predicted by the network dose calculation algorithm. Additional samples are showcased in appendix \ref{sec:appendix_a}.

\subsection{Model generalization}

In order to assess the generalization of the LSTM dose calculation engine to previously unseen patients, i.\,e., data from other patients that was not considered during training, the performance of the network was evaluated on five additional lung cancer patients. For each patient, \num{200} pencil beams with randomly selected gantry angles and isocenter shifts were prepared, and the deposited dose was calculated using MC calculations. Table \ref{tab:generalization} lists the result of comparing the MC calculations with the network estimations using $\gamma$-index analysis.

\begin{table}
	\caption{$\gamma$-index analysis on 5 different lung cancer patients}
	\centering
	\begingroup
	\setlength{\tabcolsep}{14pt}
	\renewcommand{\arraystretch}{1.4}
	\begin{tabular}{lcccc}
		\hline\hline
		&mean&std&min&max \\ [0.5ex]
		\hline
		Patient 0$^{1}$& 99.33 & 0.92 & 94.91 & 100\\
		\hline
		Patient 1& 99.10 & 0.93 & 93.01 & 99.99\\
		Patient 2& 99.01 & 1.03 & 94.96 & 100\\
		Patient 3& 99.15 & 1.00 & 94.27 & 100\\
		Patient 4$^{2}$& 97.94 & 2.27 & 85.07 & 100\\ 
		Patient 5$^{2}$& 96.34 & 4.09 & 75.79 & 99.96\\
		\hline
		\multicolumn{5}{l}{$^{1}$\scriptsize{Network has been trained on this patient.}} \\
		\multicolumn{5}{l}{$^{2}$\scriptsize{Patients with very low RSP values in lung (further discussed in section \ref{sec_discussion})}} \\
	\end{tabular}
	\endgroup
	\label{tab:generalization}
\end{table}

\subsection{Run-times}

In order to compare the run-times of incorporating the trained network as apposed to the MC algorithm, table \ref{tab:runtimes} lists the average run-times for estimating the dose for the 5 above-mentioned patients for a single pencil beam, for both methods. The MC simulations were performed with Topas with a calculation node with \num{28} virtual CPUs on an Openstack\footnote{https://www.openstack.org/} cluster. For the trained network, the run times were measured for two systems with different GPUs. Depending on the facilitated hardware, we measure average run-times of \SIrange{6}{23}{\milli\second} for the LSTM approach. Note that this run-times included the time required to send the input CT cube for each pencil beam from CPU to GPU and vice versa for the yielded dose cube. However, in applications such as adaptive radiotherapy which requires repetitive online dose estimations, the input CT cubes can be prepared and sent to the GPU in advance.  Consequently, the only relevant run-times would be the network feed forward, i.e. matrix multiplication operations run-times, reported to be \SIrange{1.5}{2.5}{\milli\second} for the two facilitated hardware stacks. The average Topas run-time was \SI{1160}{\second}, performed with $\sim\num{2.5e6}$ histories on average (see section \ref{sec:dosimetricevaluatoni}).

\begin{table}
	\caption{run-time comparison of the MC calculation vs. ANN predictions. Run times reported in parenthesis considers purely the network feed forward time consumption and does not count the time required to send each input/output from CPU to GPU and vice versa}
	\centering
	\begingroup
	\setlength{\tabcolsep}{14pt}
	\renewcommand{\arraystretch}{1.4}
	\begin{tabular}{lccc}
		\hline\hline
		&MC$^{1}$&ANN$^{2}$&ANN$^{3}$\\ [0.5ex]
		\hline
		Average run time (\si{\second})& 1159.5 & 0.023 (0.0025) & 0.006 (0.0015)\\ [1ex]
		\hline\hline
		\multicolumn{4}{l}{$^{1}$\scriptsize{28 VCPUs, 64 Gb RAM}}\\
		\multicolumn{4}{l}{$^{1}$\scriptsize{Intel Core i7-6700 3.4 GHz - Nvidia GTX 970 - 64 Gb RAM}}\\
		\multicolumn{4}{l}{$^{1}$\scriptsize{Intel Xeon W-2135 3.7 GHz - Nvidia Quadro RTX 6000 - 64 Gb RAM}}
	\end{tabular}
	\endgroup
	\label{tab:runtimes}
\end{table}
\section{Discussion}
\label{sec_discussion}

In this paper, we have demonstrated the general feasibility of proton dose calculation based on an LSTM neural network. The LSTM network correctly models the proton dose deposition characteristics in the entrance, in the Bragg peak, and in the distall fall-off region - also in heterogeneous geometries. This covers particularly examples where conventional pencil beam algorithms fail, e.\,g.\ predicting a smooth bi-modal Bragg peak behind interfaces. 

In comparison to RNN networks, LSTM networks proved particularly suited for this task, especially in heterogeneous geometries. This was also reflected in the training behavior, where the RNNs exhibited more pronounced fluctuations in MSE loss (compare figure \ref{fig:lossPlotBox}).

Using phantom and lung patient cases, we have observed very good agreement for individual pencil beams with an initial energy of \SI{104.25}{\mega\electronvolt} at run-times of \SIrange{6}{23}{\milli\second} per pencil beam. While the $\gamma$-index pass rates for patients \numrange{1}{3} was $>$\SI{99}{\percent}, the $\gamma$-index pass rates for patients \num{4} and \num{5} ranged between \SI{96}{\percent} and \SI{98}{\percent}. This slight decline was attributed to very low \si{RSP} values in lung which could not be discriminated against air volumes penetrated by the beam before entering the patient. This phenomenon originated from beam angles in the training set where the beam enters and exits the patient arms before impinging on the chest (see figures \ref{fig:LSTMpatientShowcaseApp_a} and \ref{fig:LSTMpatientShowcaseApp_p}). Even though these beam orientations would be probably excluded from clinical considerations, we decided to have them included as challenging test scenario for the networks.

Based on the approach to study dose calculation accuracy for an individual energy, we were able to show the generalization of our algorithm to patient cases that were not considered during LSTM training. In order to implement a dose calculation for an entire treatment plan, however, additional networks need to be trained for different energies. Alternatively, and conceptually more appealing, it may be possible to train a network that is able to generalize also over different initial energies.

Of course, the run time benefits of several orders of magnitude over MC simulations as shown in table \ref{tab:runtimes} will not manifest in the same way for clinical treatment plans comprised of several thousands of pencil beams. Here, MC simulations can save substantially because the geometry will only be initialized once for the entire simulation. Furthermore it will be possible to reduce the number of histories per pencil beam to achieve sufficient statistical certainty over the entire treatment plan for a simple dose recalculation. For the computation of a dose influence matrix which is needed for dose optimization, however, the MC runtime reductions will be more moderate. On the other side, it will be possible to further accelerate LSTM-based dose calculation through dedicated deep learning hardware and leveraging the embarrassingly parallel nature of the problem. And, as previously indicated, the transfer times between CPU and GPU will only be necessary once per patient for LSTM network dose calculations. In our case this made up \SI{75}{\percent} of the run time for the faster GPU hardware.

Moreover, this study was concentrated on the ability to estimate dose in heterogeneous geometries, and no effort was made in improving the model efficiency. In this regard, there exist various model compression techniques, e.g. pruning, quantization, and tensor decomposition methods (achieving low-rank structures in the weight matrices) \cite{Grachev_2019, pmlr-v70-yang17e, ye2017learning}, which can lower the number of parameters in fully connected layers substantially \cite{Yang_2015_ICCV, han2015deep}. The efficiency of the model can be further enhanced through fine-tuning of the model architecture. This study parameterized the size in longitudinal direction as a fixed hyper-parameter (parameter $l$, see section \ref{sec:parameterization}). While the range of mono-energetic protons are more or less fixed in a homogeneous geometry, it can vary substantially when they travel through wide cavities such as the Lung. This issue coerces us to train the model with very long sequences, to encompass all the potential pencil beam ranges. However, the LSTM models can be designed in what is referred to as \emph{sequence to sequence learning}, which can accept a variable length input and outcome with a variable length output, incorporated effectively in Machine Translation problems \cite{NIPS2014_5346}. Utilization of such a model can restrict the number of matrix multiplication operations accustomed to the plan, resulting in even faster estimations. In a different approach, one could also incorporate Autoencoders \cite{baldi2012autoencoders}, as a back-end to the model, compressing the input CT to a latent feature space, leading to a reduction in number of input parameters.

The proposed dose estimation approach has not been exploited so far to the best of our knowledge, and we intend to explore this approach in many aspects. We see possible applications in photon dose calculation as well as in heavier ions (Carbon, Oxygen, Helium) dose calculation in an attempt to estimate biologically effective dose distributions.

\section{Conclusion}
\label{sec_conclusion}

In this paper, we have investigated the role of two different neural network architectures for proton dose calculation, i.\,e., an RNN and an LSTM network. For individual pencil beams on varying heterogeneous phantom geometries, the average $\gamma$-index pass rate ([\SI{0.5}{\percent} , \SI{1}{mm}]) was \SI{97.9}{\percent} for the RNN and \SI{99.3}{\percent} for the LSTM network. The LSTM network was further evaluated on a highly heterogeneous lung case where we observed an average $\gamma$-index pass rate of \SI{99.3}{\percent} ([\SI{0.5}{\percent} , \SI{2}{mm}]). Average LSTM network run-times ranged between \SIrange{6}{23}{\milli\second}.

Our results indicate that LSTM networks are well suited for particle therapy dose calculation tasks. 

\section{Acknowledgements}

The authors thank Lucas Burigo for providing the TOPAS Monte Carlo interface for matRad.

\bibliographystyle{unsrt}  


\appendix
\pagebreak

\section{Additional output}
\label{sec:appendix_a}

In this section, we illustrate the performance of the trained network on additional challenging test samples, in which the inputs exhibit noticeable heterogeneities. These samples are extracted from the test data set of the patient \num{0}, patient \num{3}, and patient \num{5}.

\begin{figure}[htb]
	\begin{center}
		\centering
		\subfigure[$\gamma$-index pass rate = \SI{96.4}{\percent}]{\includegraphics[width=0.45\linewidth]{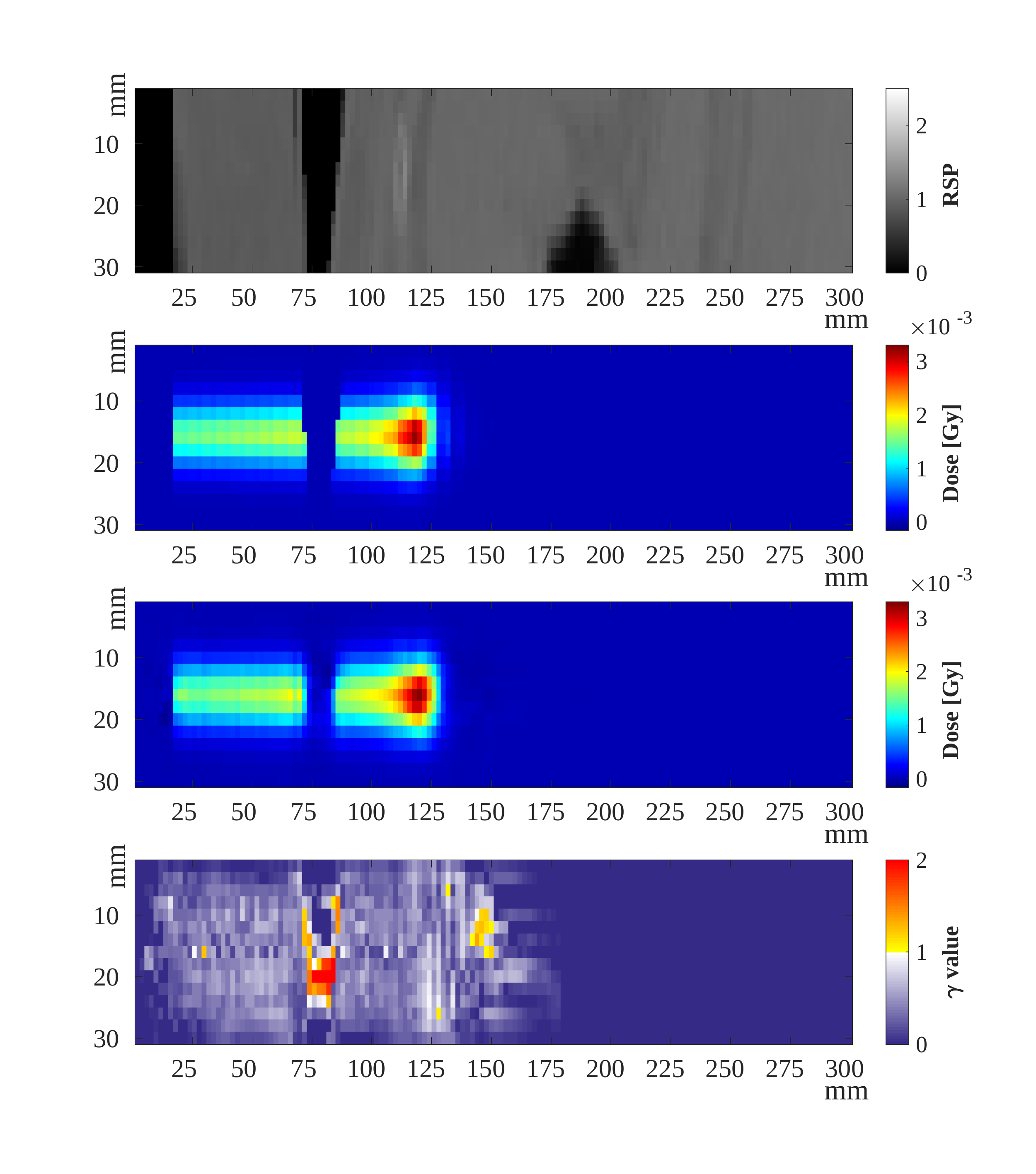}\label{fig:LSTMpatientShowcaseApp_a}}\hfill
		\subfigure[$\gamma$-index pass rate = \SI{99.1}{\percent}]{\includegraphics[width=0.45\linewidth]{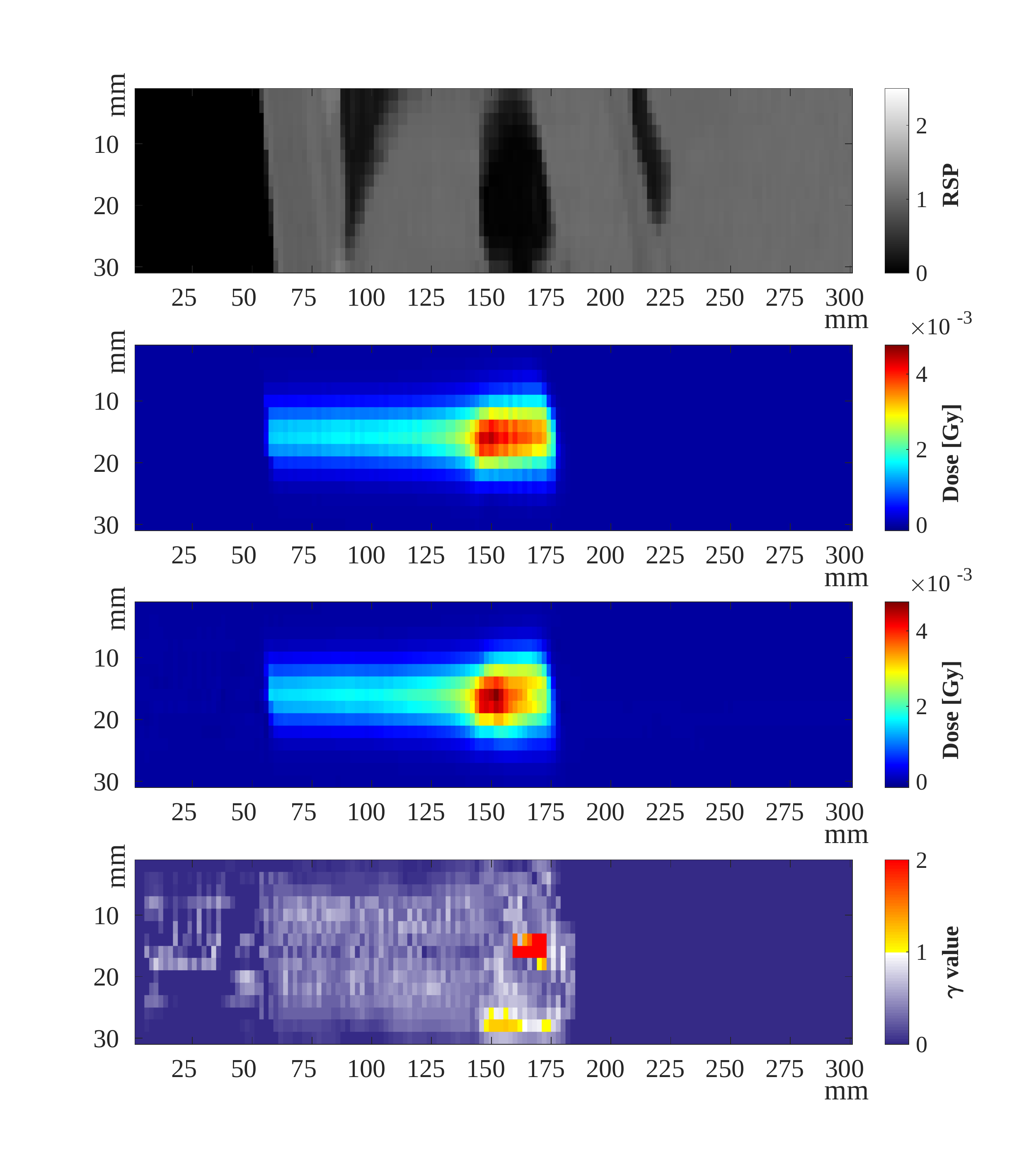}\label{fig:LSTMpatientShowcaseApp_b}}\\
		\subfigure[$\gamma$-index pass rate = \SI{97.5}{\percent}]{\includegraphics[width=0.45\linewidth]{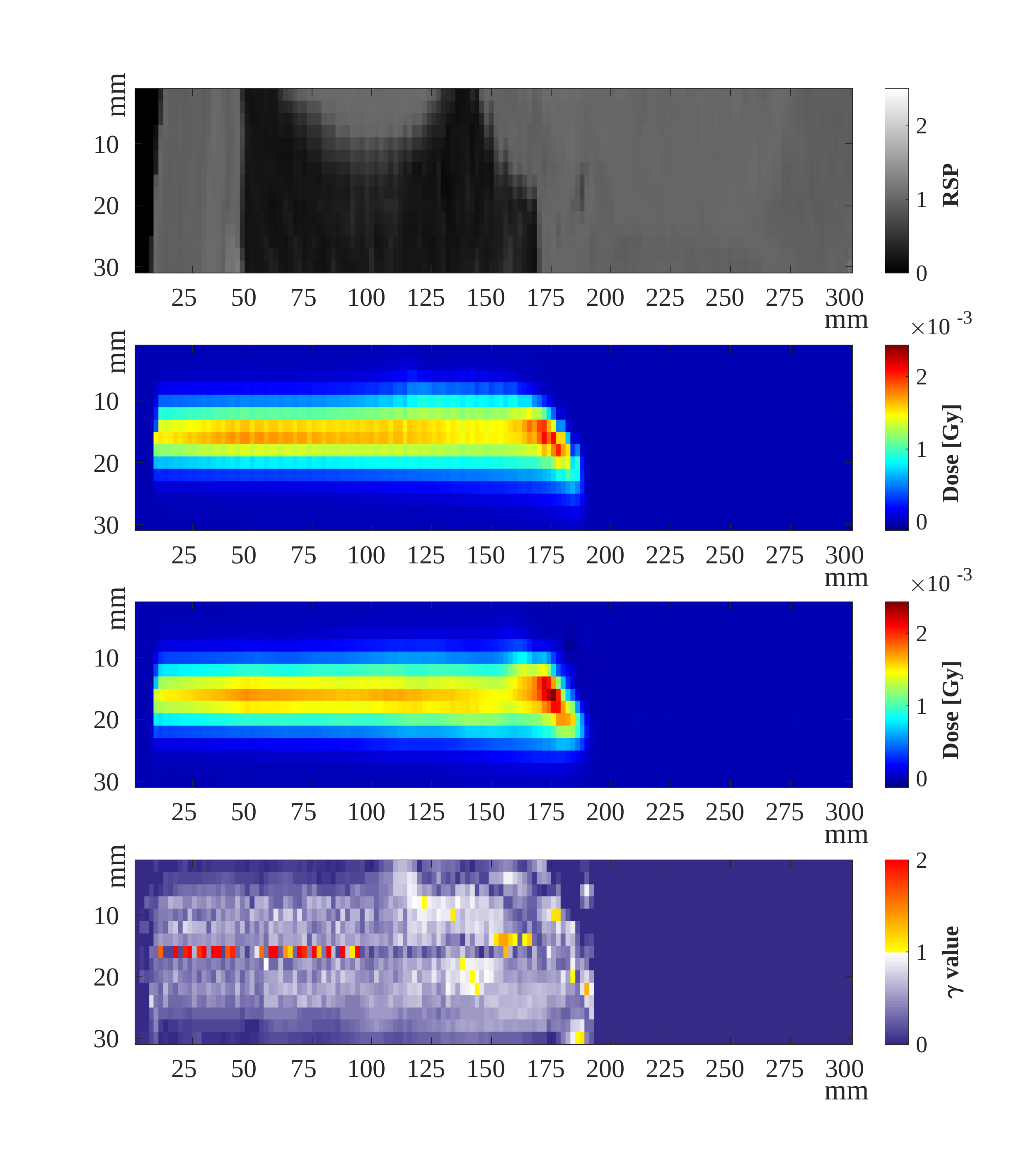}\label{fig:LSTMpatientShowcaseApp_c}}\hfill
		\subfigure[$\gamma$-index pass rate = \SI{97.6}{\percent}]{\includegraphics[width=0.45\linewidth]{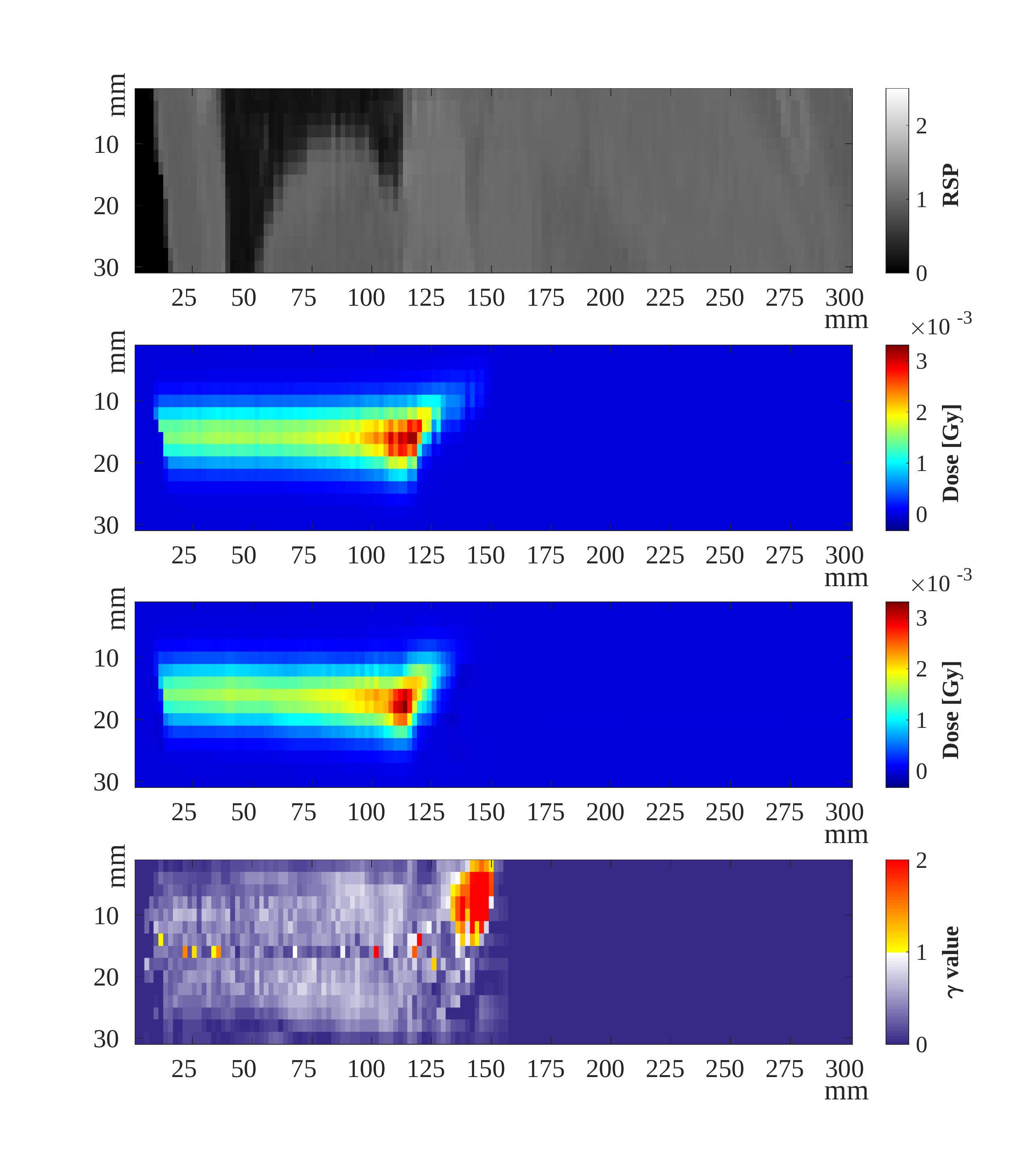}\label{fig:LSTMpatientShowcaseApp_d}}\\
		\caption{Dose estimation results for \num{4} test data from patient \num{0} ($E = $ \SI{104.25}{\mega\electronvolt}, $\gamma$-index criteria $ = $ [\SI{0.5}{\percent} , \SI{2}{mm}]). The Subfigurs follow the layout outlined in figure \ref{fig:LSTMpatientShowcase}. Note that sample (a) has an air gap between the patient's arm and chest (discussed in section \ref{sec_discussion}).}
	\end{center}
	\label{fig:LSTMpatientShowcaseApp_1}
\end{figure}

\begin{figure}[htb]
	\begin{center}
		\centering
		\subfigure[$\gamma$-index pass rate = \SI{97.8}{\percent}]{\includegraphics[width=0.45\linewidth]{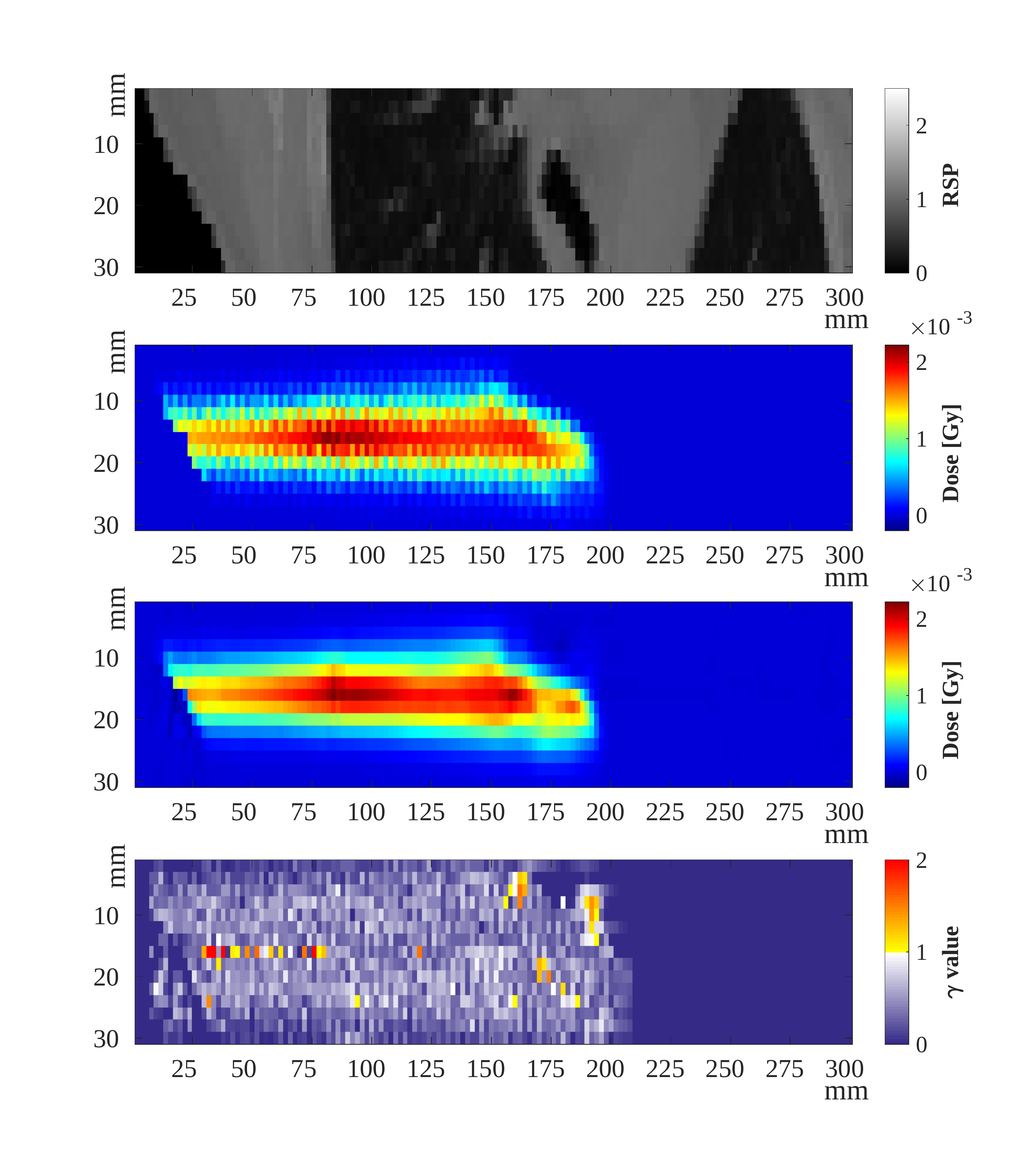}\label{fig:LSTMpatientShowcaseApp_e}}\hfill
		\subfigure[$\gamma$-index pass rate = \SI{97.1}{\percent}]{\includegraphics[width=0.45\linewidth]{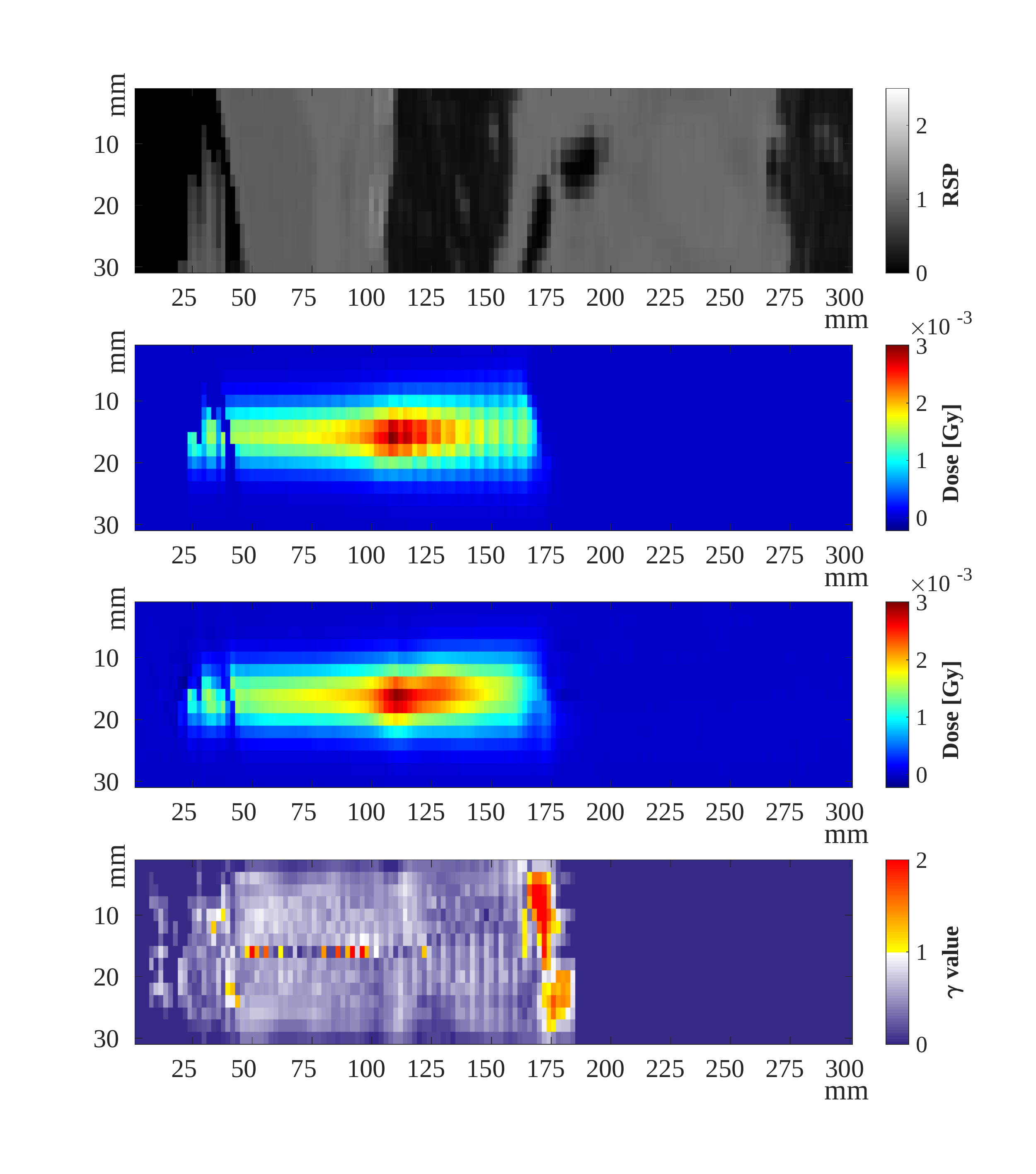}\label{fig:LSTMpatientShowcaseApp_f}}\\
		\subfigure[$\gamma$-index pass rate = \SI{98.3}{\percent}]{\includegraphics[width=0.45\linewidth]{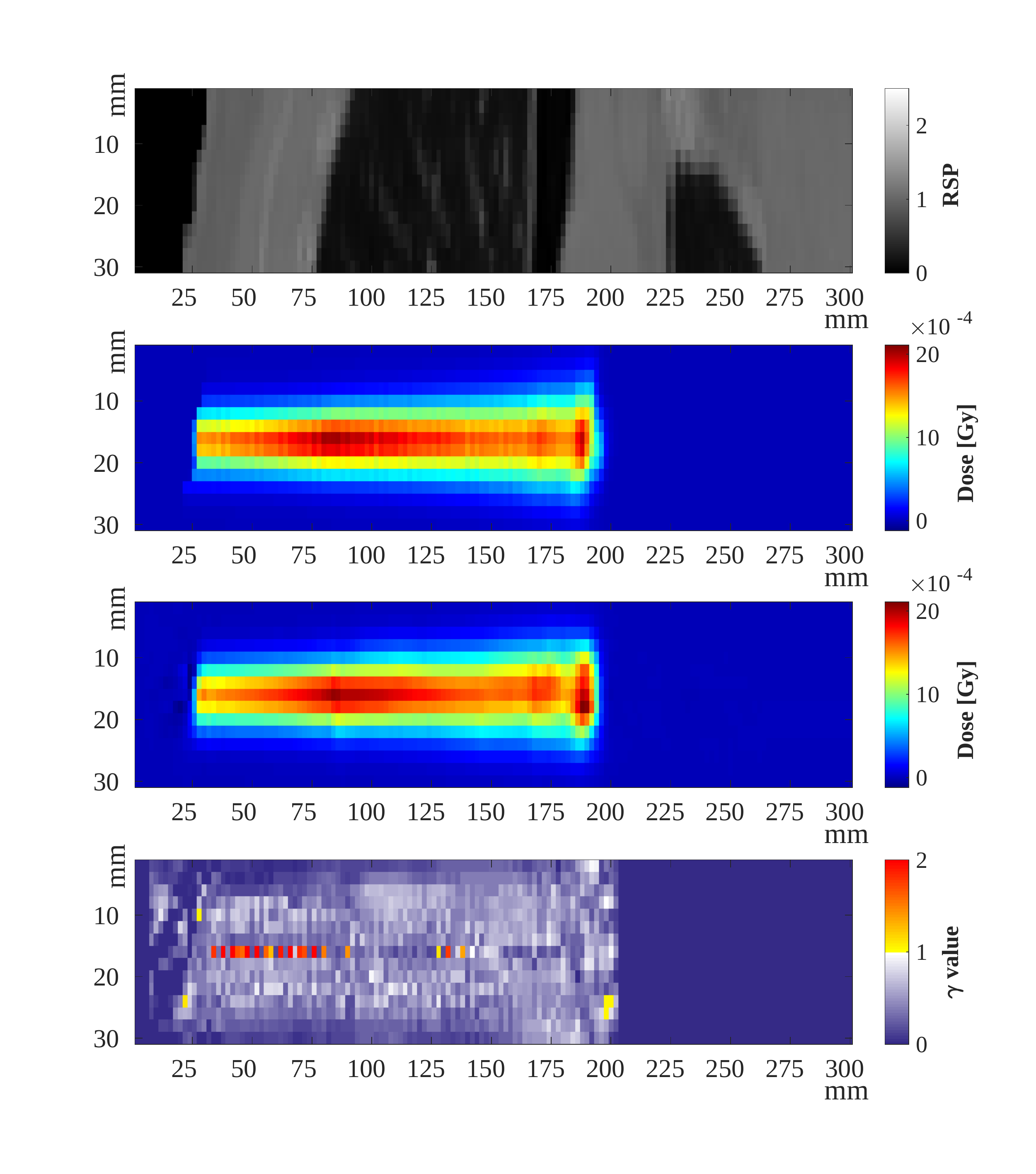}\label{fig:LSTMpatientShowcaseApp_g}}\hfill
		\subfigure[$\gamma$-index pass rate = \SI{98.7}{\percent}]{\includegraphics[width=0.45\linewidth]{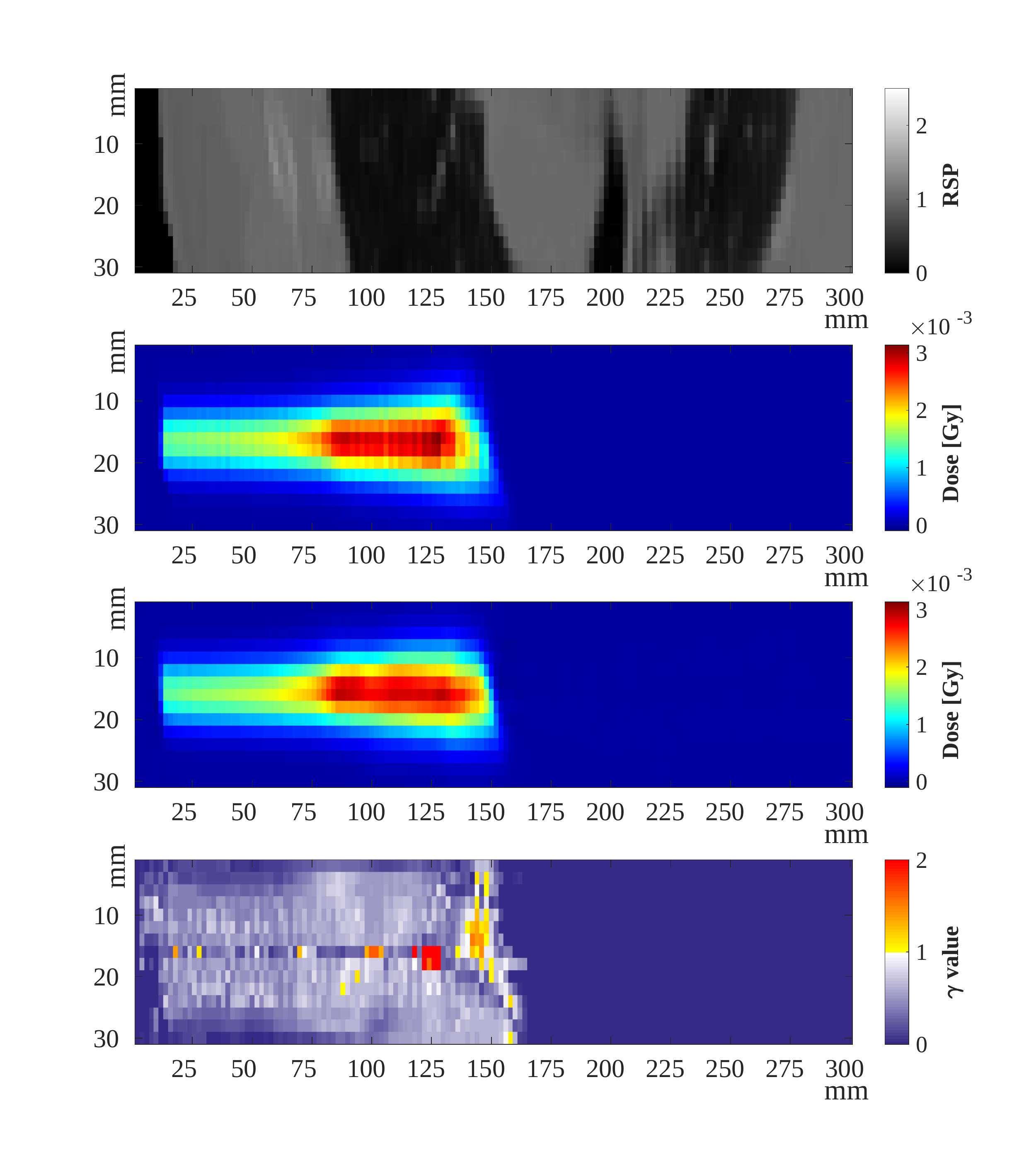}\label{fig:LSTMpatientShowcaseApp_h}}\\
		\label{fig:LSTMpatientShowcaseApp_2}
		\caption{Dose estimation results for \num{4} test data from patient \num{0} ($E = $ \SI{104.25}{\mega\electronvolt}, $\gamma$-index criteria $ = $ [\SI{0.5}{\percent} , \SI{2}{mm}]). The Subfigurs follow the layout outlined in figure \ref{fig:LSTMpatientShowcase}. Note that the aliasing effect in sample (a) is due to the cube extraction interpolation of oblique gantry angles.}
	\end{center}
\end{figure}

\begin{figure}[htb]
	\begin{center}
		\centering
		\subfigure[$\gamma$-index pass rate = \SI{98.1}{\percent}]{\includegraphics[width=0.45\linewidth]{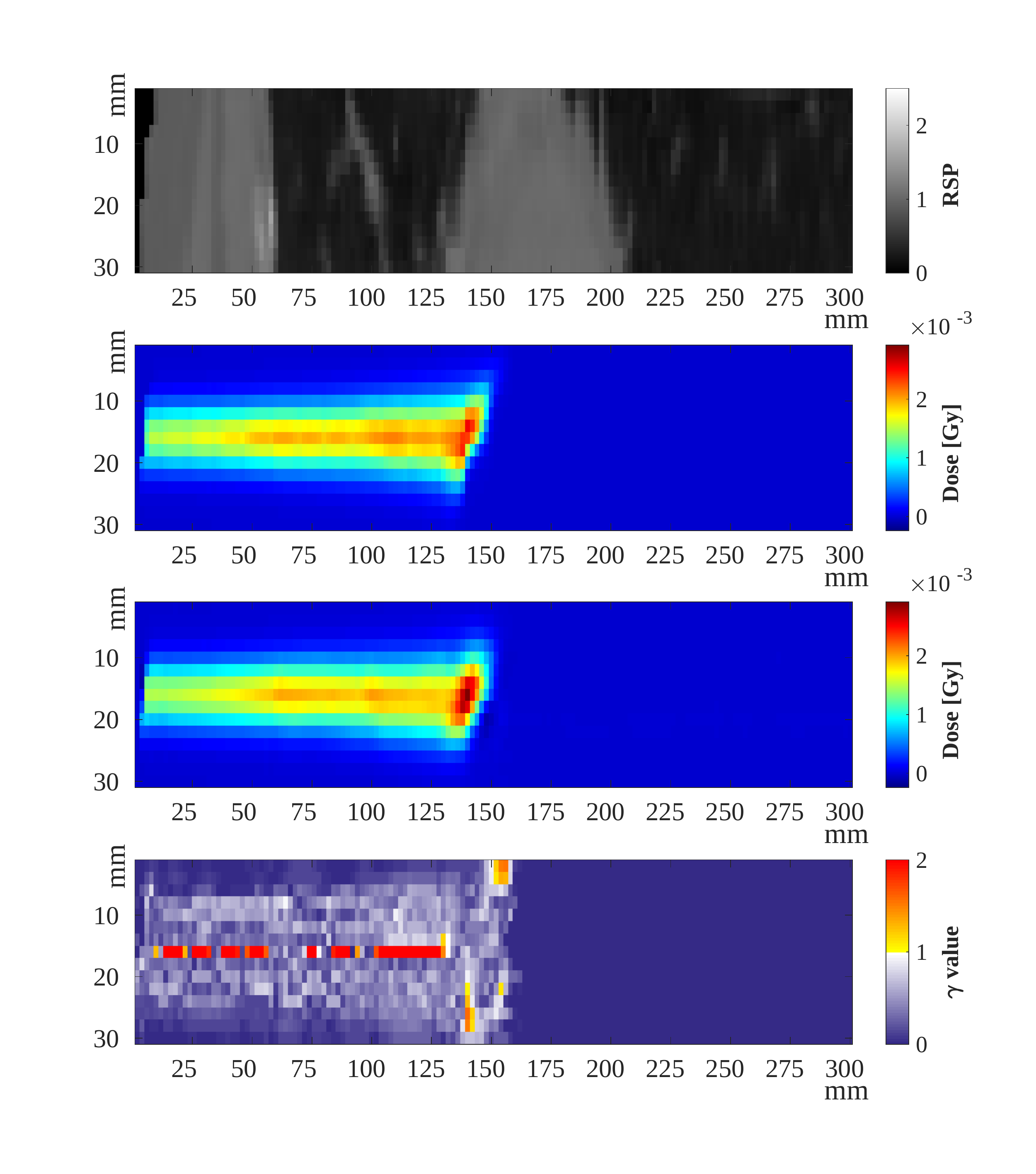}\label{fig:LSTMpatientShowcaseApp_i}}\hfill
		\subfigure[$\gamma$-index pass rate = \SI{98.4}{\percent}]{\includegraphics[width=0.45\linewidth]{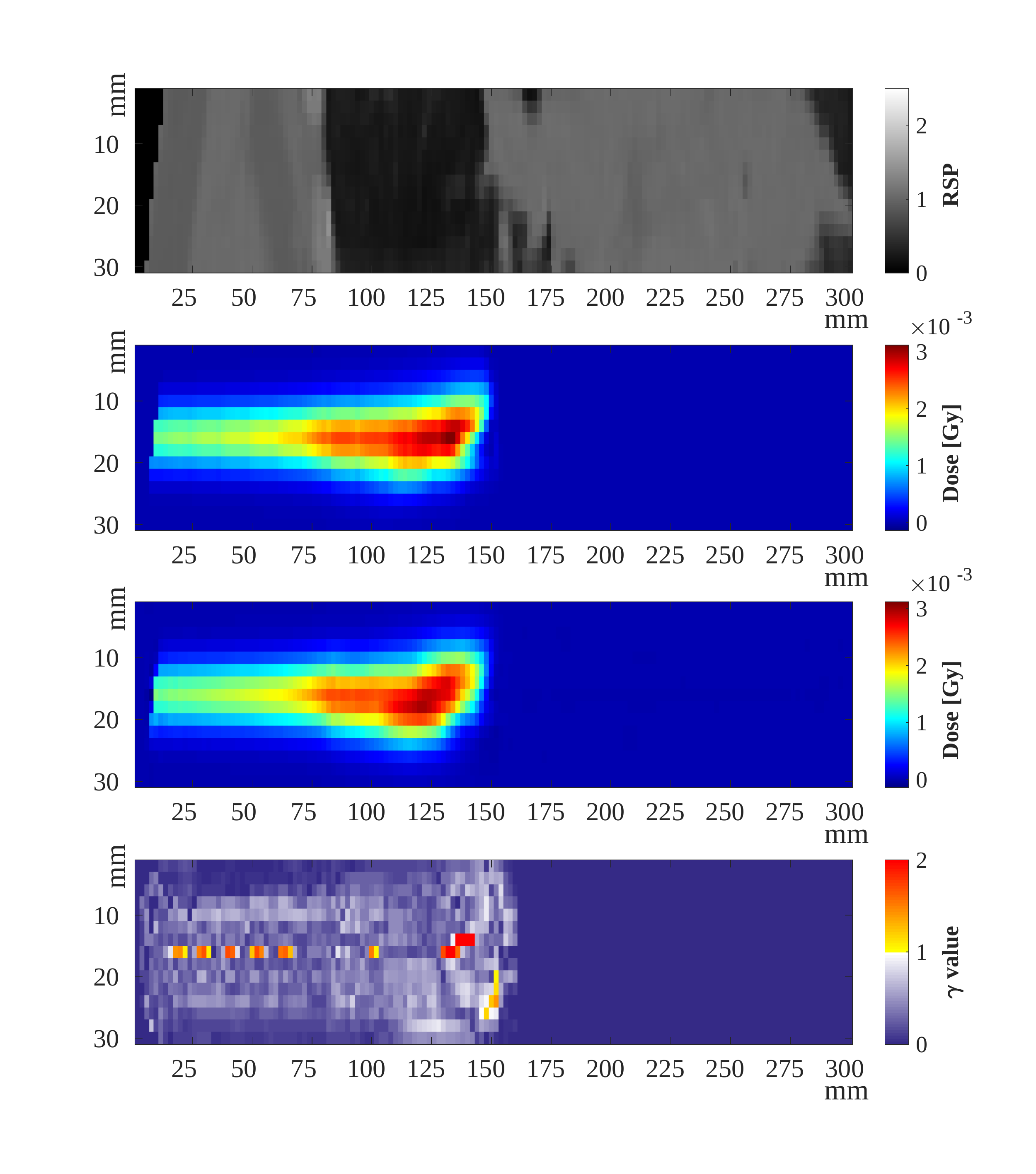}\label{fig:LSTMpatientShowcaseApp_j}}\\
		\subfigure[$\gamma$-index pass rate = \SI{98.6}{\percent}]{\includegraphics[width=0.45\linewidth]{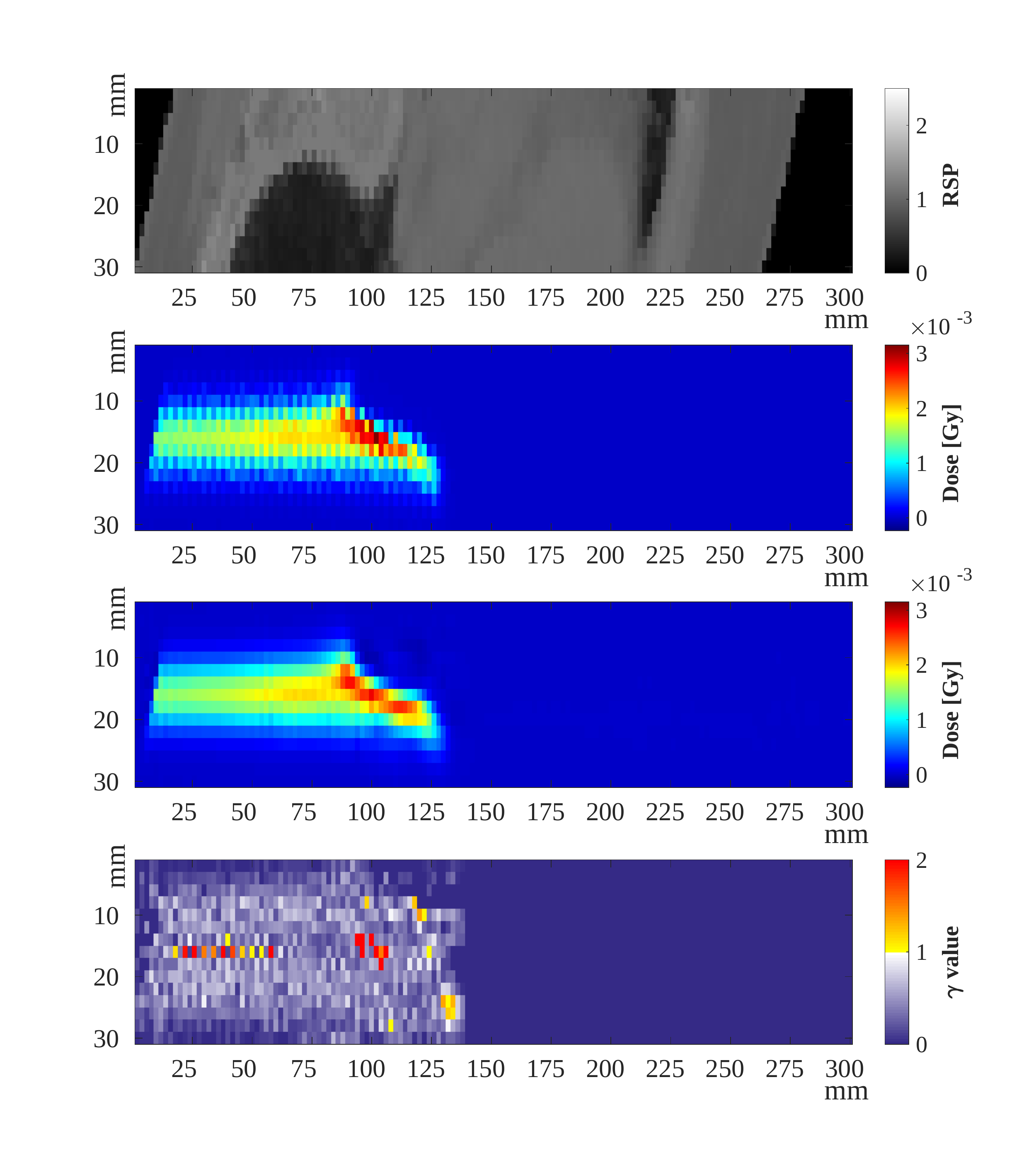}\label{fig:LSTMpatientShowcaseApp_k}}\hfill
		\subfigure[$\gamma$-index pass rate = \SI{98}{\percent}]{\includegraphics[width=0.45\linewidth]{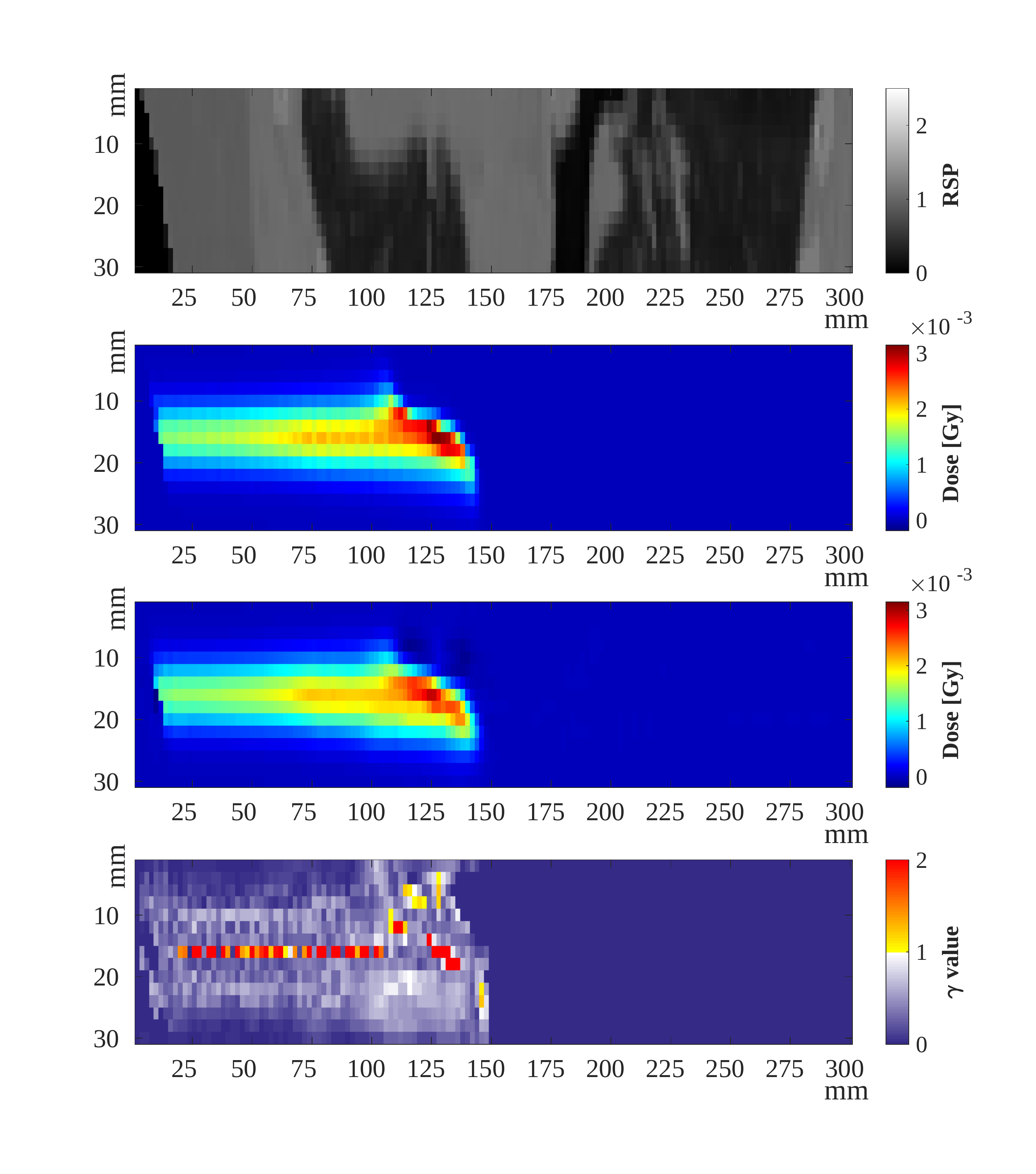}\label{fig:LSTMpatientShowcaseApp_l}}\\
		\label{fig:LSTMpatientShowcaseApp_3}
		\caption{Dose estimation results for \num{4} test data from patient \num{3} ($E = $ \SI{104.25}{\mega\electronvolt}, $\gamma$-index criteria $ = $ [\SI{0.5}{\percent} , \SI{2}{mm}]). The Subfigurs follow the layout outlined in figure \ref{fig:LSTMpatientShowcase}.}
	\end{center}
\end{figure}

\begin{figure}[htb]
	\begin{center}
		\centering
		\subfigure[$\gamma$-index pass rate = \SI{98.2}{\percent}]{\includegraphics[width=0.45\linewidth]{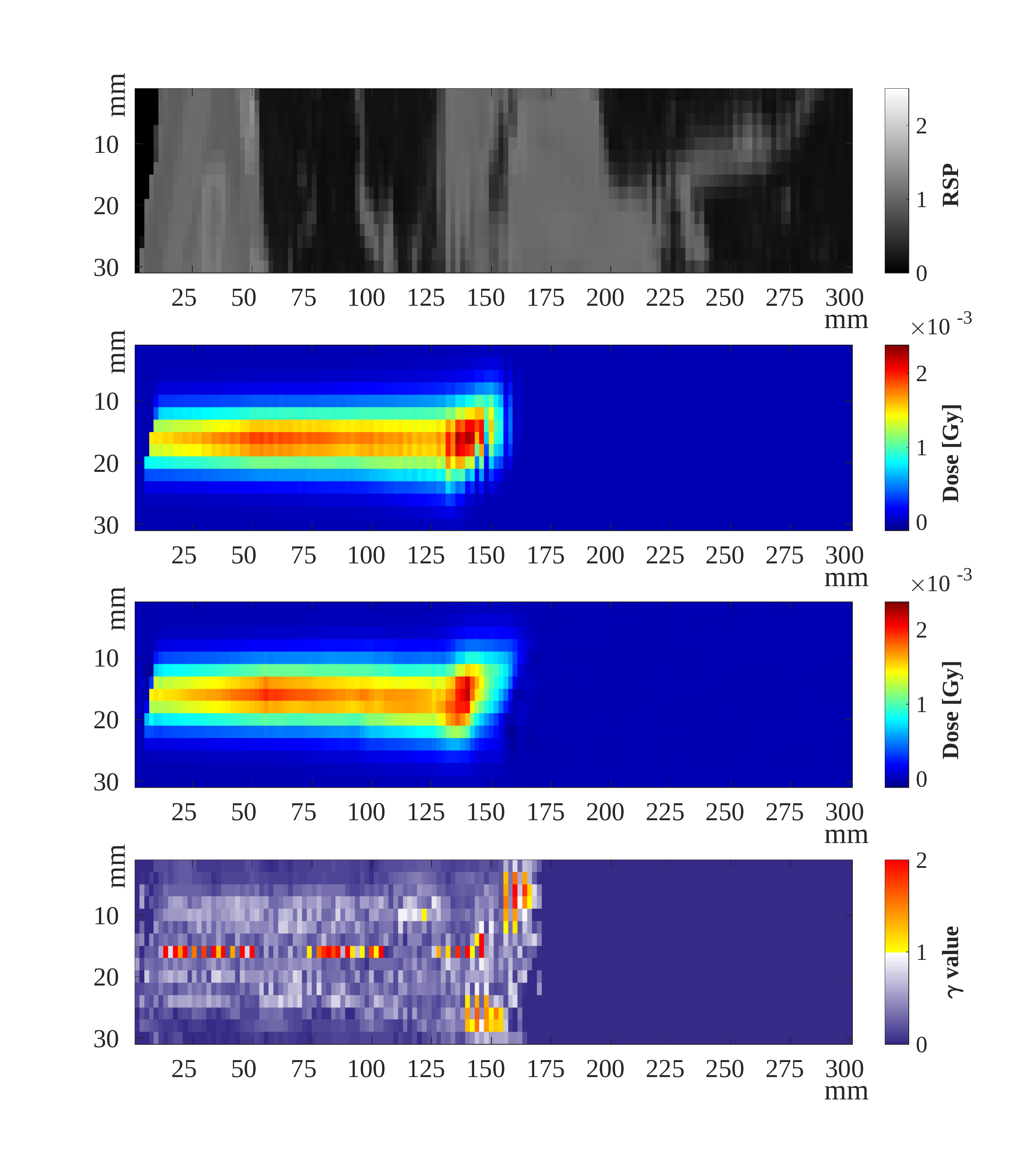}\label{fig:LSTMpatientShowcaseApp_m}}\hfill
		\subfigure[$\gamma$-index pass rate = \SI{93.2}{\percent}]{\includegraphics[width=0.45\linewidth]{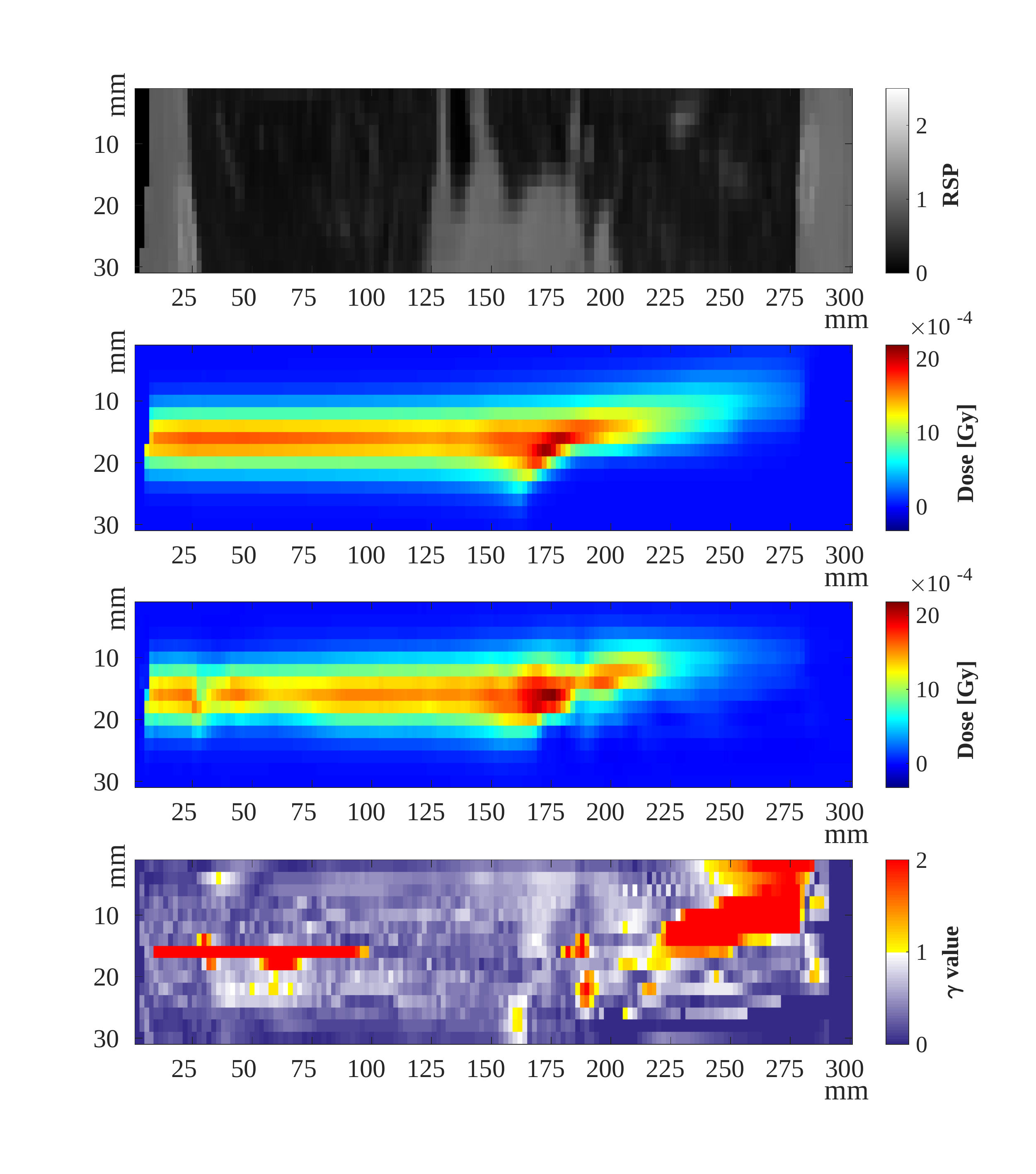}\label{fig:LSTMpatientShowcaseApp_n}}\\
		\subfigure[$\gamma$-index pass rate = \SI{97.5}{\percent}]{\includegraphics[width=0.45\linewidth]{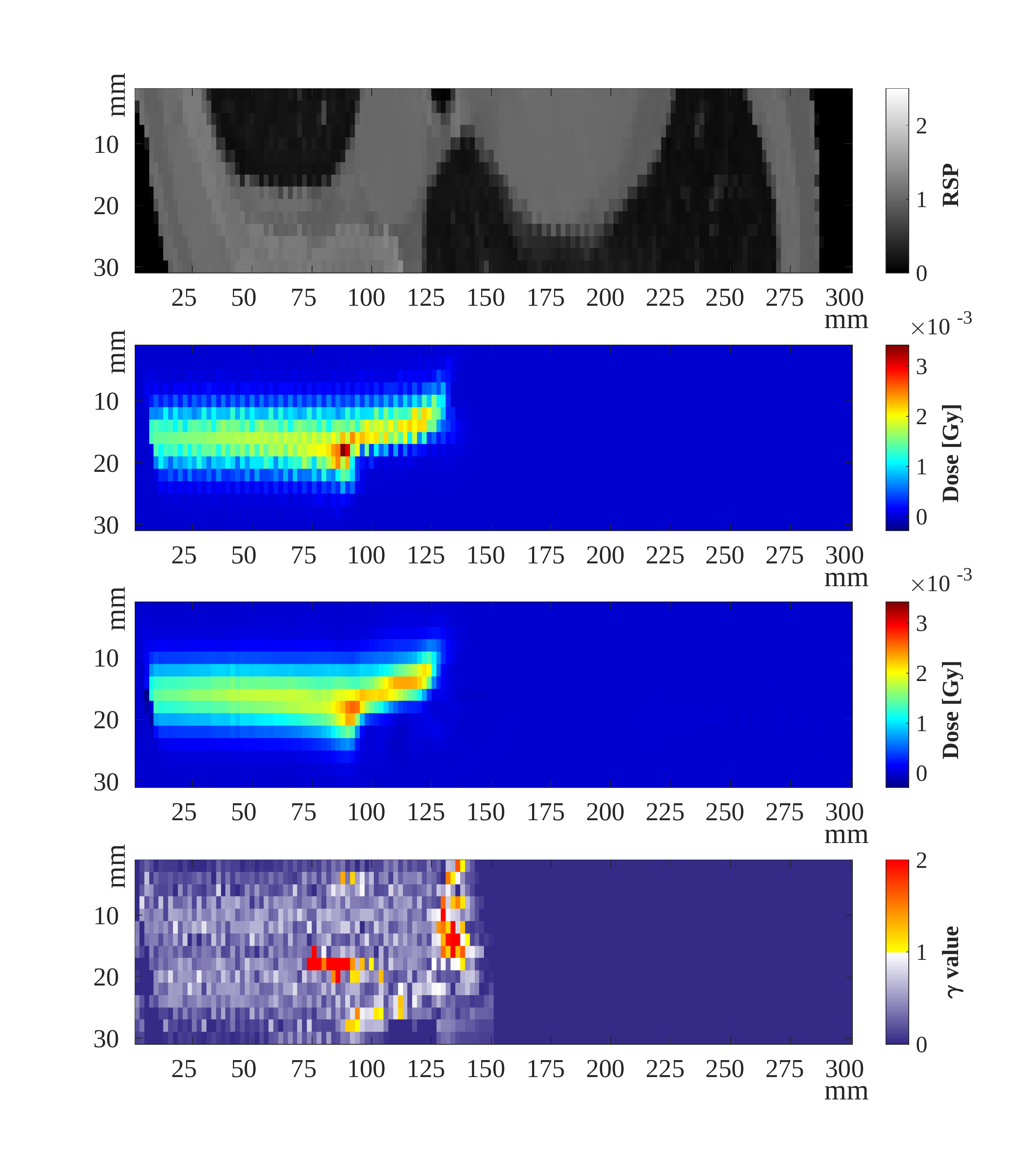}\label{fig:LSTMpatientShowcaseApp_o}}\hfill
		\subfigure[$\gamma$-index pass rate = \SI{91.4}{\percent}]{\includegraphics[width=0.45\linewidth]{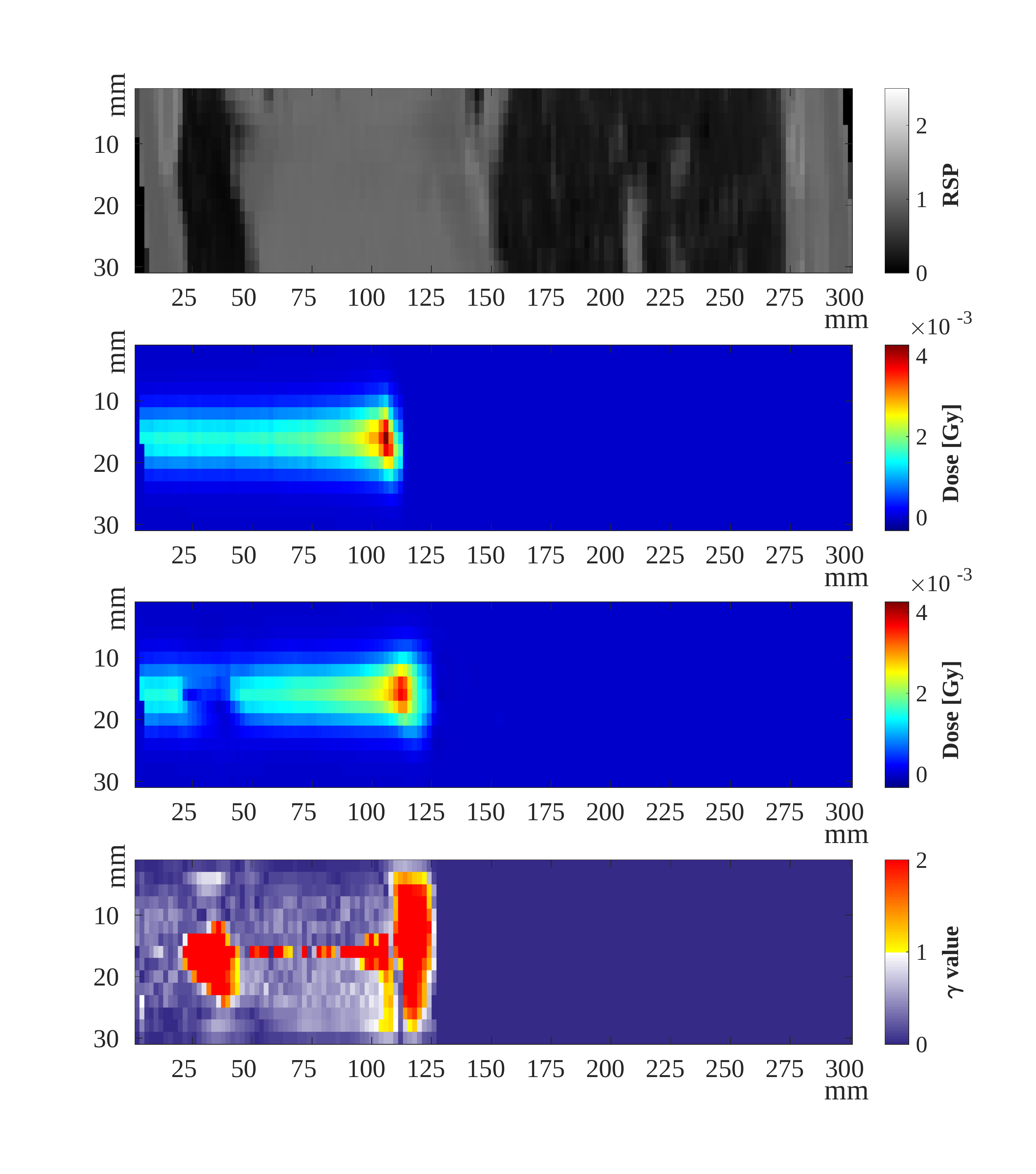}\label{fig:LSTMpatientShowcaseApp_p}}\\
		\label{fig:LSTMpatientShowcaseApp_4}
		\caption{Dose estimation results for \num{4} test data from patient \num{5} ($E = $ \SI{104.25}{\mega\electronvolt}, $\gamma$-index criteria $ = $ [\SI{0.5}{\percent} , \SI{2}{mm}]). The Subfigurs follow the layout outlined in figure \ref{fig:LSTMpatientShowcase}. Note that the network fails to distinguish between the lung and the air cavity in sample (d) due to the very low \si{RSP} value of the lung.}
	\end{center}
\end{figure}

%
\end{document}